\documentclass[namedreferences]{SolarPhysics}
\usepackage[optionalrh,natbib]{spr-sola-addons}
\usepackage{graphicx}        
\usepackage{color}           
\usepackage{url}             

\renewcommand{\vec}[1]{ {\mathbf #1} }

\newcommand{\grad}{ {\bf \nabla } }
\newcommand{\curl}{ {\bf \nabla} \times}
\newcommand{\bb}{ \vec B}
\newcommand{\jj}{ \vec j}
\newcommand{\bg}{\mbox{\boldmath$g$}}
\newcommand{\bxi}{\mbox{\boldmath $\xi$}}

\begin{document}

\begin{article}

\begin{opening}

\title{3D Numerical Simulations of $f$-Mode Propagation Through
  Magnetic Flux Tubes}

\author{K.~\surname{Daiffallah}$^{1}$\sep
        T.~\surname{Abdelatif}$^{1}$\sep
        A.~\surname{Bendib}$^{2}$\sep
R.~\surname{Cameron}$^{3}$\sep
L.~\surname{Gizon}$^{3}$     
       }
\runningauthor{K.Daiffallah \textit{et al}.}
\runningtitle{$f$-Mode Interactions with Magnetic Flux Tubes}

   \institute{$^{1}$ Observatory of Algiers, CRAAG, Algiers, Algeria, 
                     email: \url{k.daiffallah@craag.dz} \\ 
              $^{2}$ University of Sciences and Technology (USTHB), Faculty of Physics, Algiers, Algeria
                    \\
$^{3}$ Max-Planck-Institut f\"{u}r Sonnensystemforschung, Katlenburg-Lindau, Germany
\\
             }

\begin{abstract}
Three-dimensional numerical simulations have been used to study the scattering of a surface-gravity wave packet by vertical magnetic flux tubes, with radii from 200 km to 3 Mm, embedded in stratified polytropic atmosphere. The scattered wave was found to consist primarily of
$m=0$ (axisymmetric) and $m=1$ modes. It was found that the ratio of
the amplitude of these two modes is strongly dependant on the radius of the flux tube: The kink mode is the dominant mode excited in tubes with a small radius while the sausage mode is dominant for large tubes. Simulations of this type provide a simple, efficient and robust way to start understanding the seismic signature of flux tubes, which have recently began to be observed.
\end{abstract} 

\keywords{Helioseismology, Direct Modeling; Waves,
  Magnetohydrodynamic; Magnetic Fields }
\end{opening}

\section{Introduction}
     \label{S-Introduction} 
Magnetic flux tubes couple the different layers of the solar
atmosphere, and waves propagating along the magnetic-field lines are a
possible source of the mechanical-energy transport required to heat
the chromosphere and corona. The waves also provide a helioseismic
means of constraining the properties of the magnetic tubes which is
complementary to observational studies with SOHO/MDI data
\citep{Duvall06} and \textit{Hinode} \citep{Fujimura09}. In the latter work the
authors found that the observed Poynting flux is substantial when
compared to the flux required to heat the chromosphere and
corona. Unfortunately with observations at a single height they were
unable to identify what mixture of $m=0$, axisymmetric sausage-modes,
and $m=1, $ kink-modes, they were observing. Theoretical studies of the interaction of flux tubes and waves have been performed in the past, for example \cite{Roberts79},  \cite{Wilson80}, \cite{Spruit81}, \cite{Spruit82}, \cite{Knolker89}, \cite{Solanki93}, 
\cite{Bogdan96},
\cite{Hasan99}, \cite{Tirry00}, \cite{Gizon06}, \cite{Jain08}, \cite{Hanasoge08a},
\cite{Hanasoge09}.
It is now well established that slender magnetic flux tubes permit the
propagation of the two basic types of magnetohydrodynamic waves: The
longitudinal tube waves (sausage modes) with azimuthal wave number $m$
= 0, which are axisymmetric, excited by pressure fluctuations. The
transversal tube waves (kink modes) with $m$ =$\pm$ 1, which are
non-axisymmetric and describe the incompressible undulations of the
tube, supported by magnetic tension. However, in these various
papers, simplifying assumptions were used, such as the thin-flux-tube
approximation, isothermal or unstratified atmospheres, and the neglect
of higher-$m$ modes. 
\cite{Hanasoge08a}
included a stratified atmosphere and used the thin-flux-tube approximation  to
evaluate the scattering matrix associated with an $f$-mode wave
interacting with a flux tube. The thin-flux-tube approximation works
when the radius of the tube is smaller than it scale height which was $260$ km. For this 
reason the largest tube that they considered was 160 km. Numerical treatment of flux tubes of 
arbitrary size is straightforward and robust. The approach is already yielding interesting 
results in the context of sunspots, see {\it e.g.} \cite{Cameron07}, 
\cite{Hanasoge08},  \cite{Khomenko08}, \cite{Cameron08}, \cite{Khomenko09}.

In this paper we use numerical simulations to explore the response of vertical, small scale
flux-tubes with radii between 200 km and 3 Mm. The paper is structured as
follows: In Section 2 we  present the background model and the
equations describing the wave propagation. In Section 3 we present the
results of a series of convergence tests. These are essential since we
are considering tubes with small radii. In Section 4 we study the scattered wave field, Section 5 is devoted to mode conversion, before discussing
the results in Section 6.

\section{Simulation Model} 
      \label{S-Simulation model}

\begin{figure}   
\centering
\includegraphics[width=0.33\textwidth]{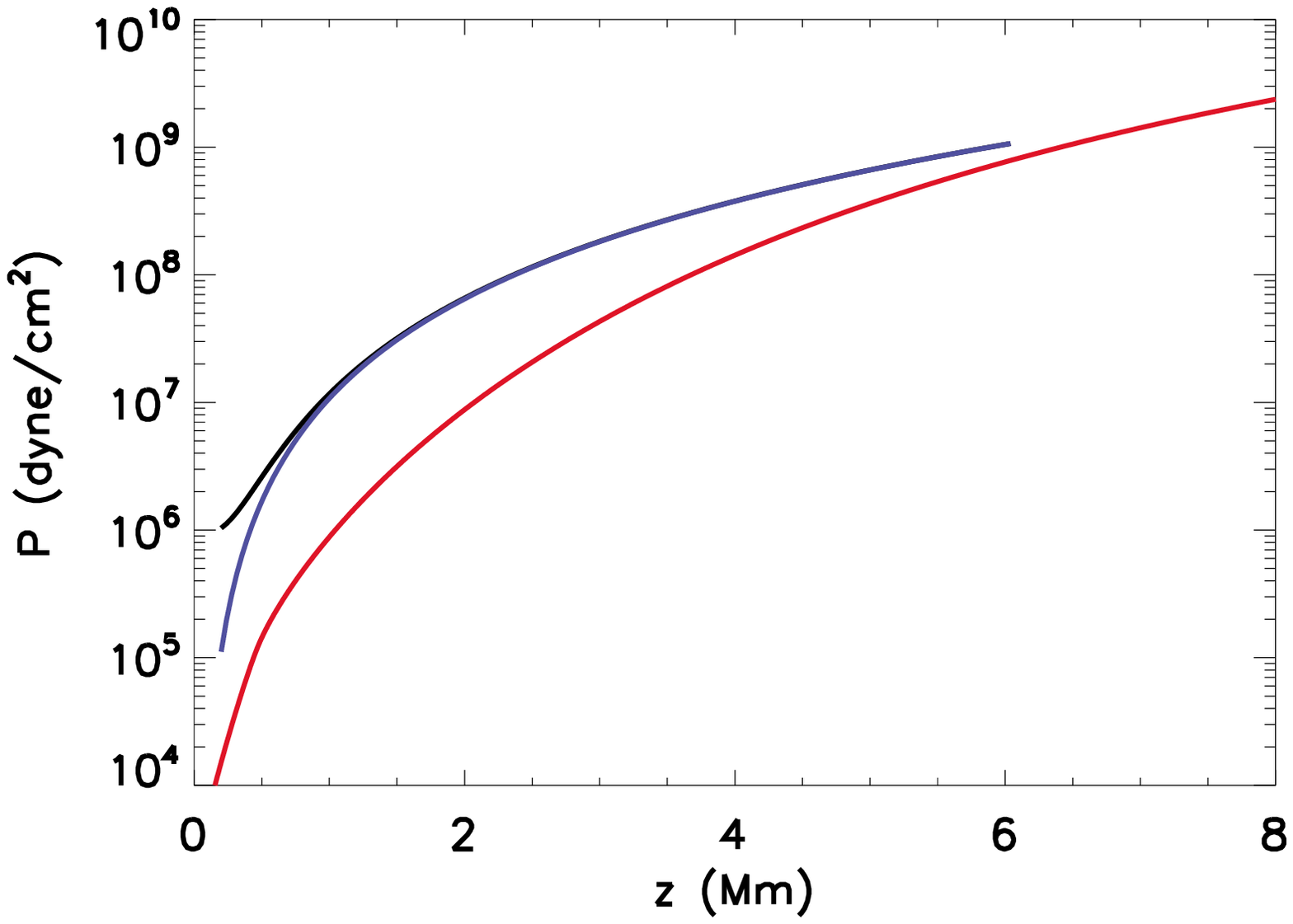}
\includegraphics[width=0.33\textwidth]{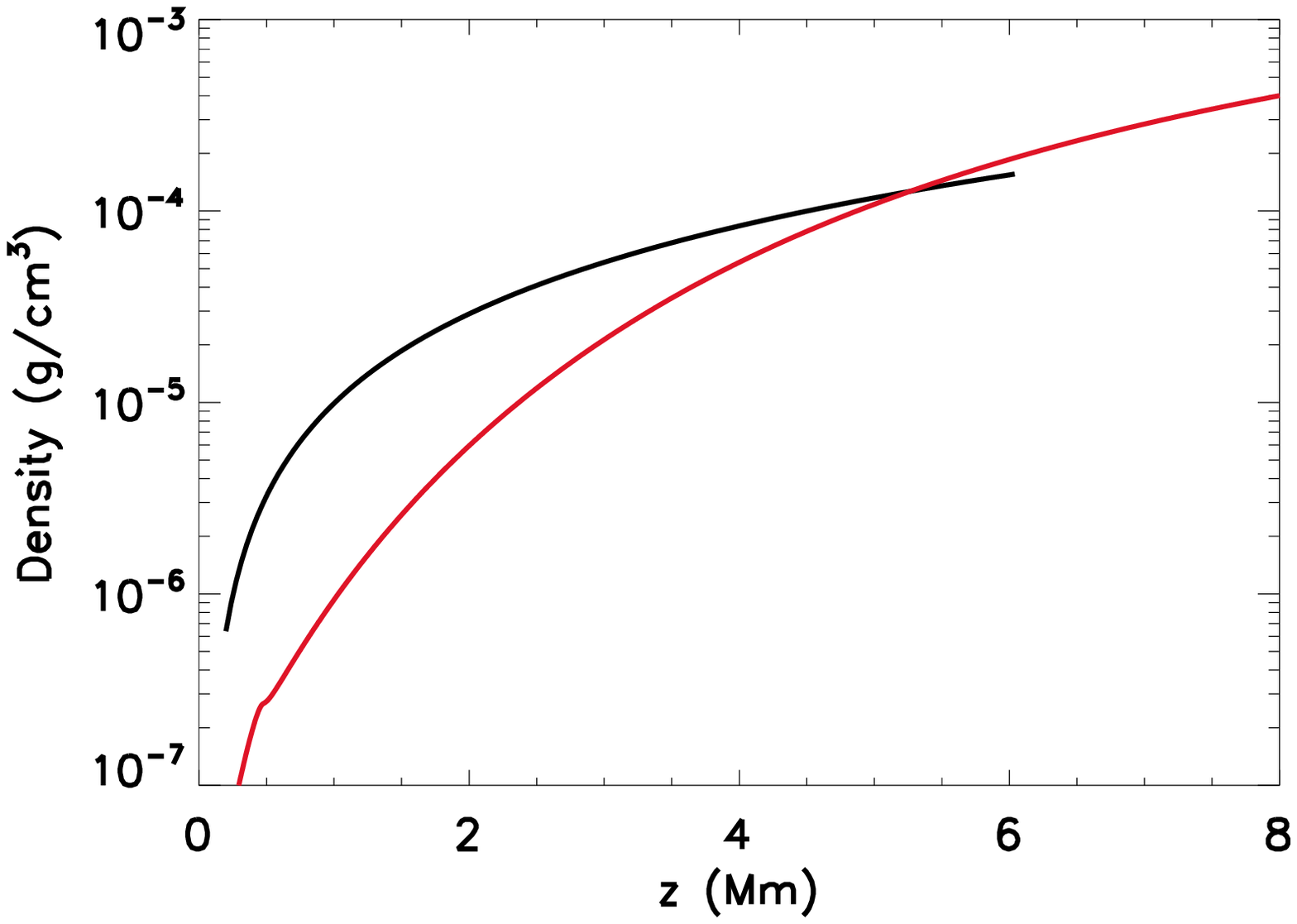}
\includegraphics[width=0.32\textwidth]{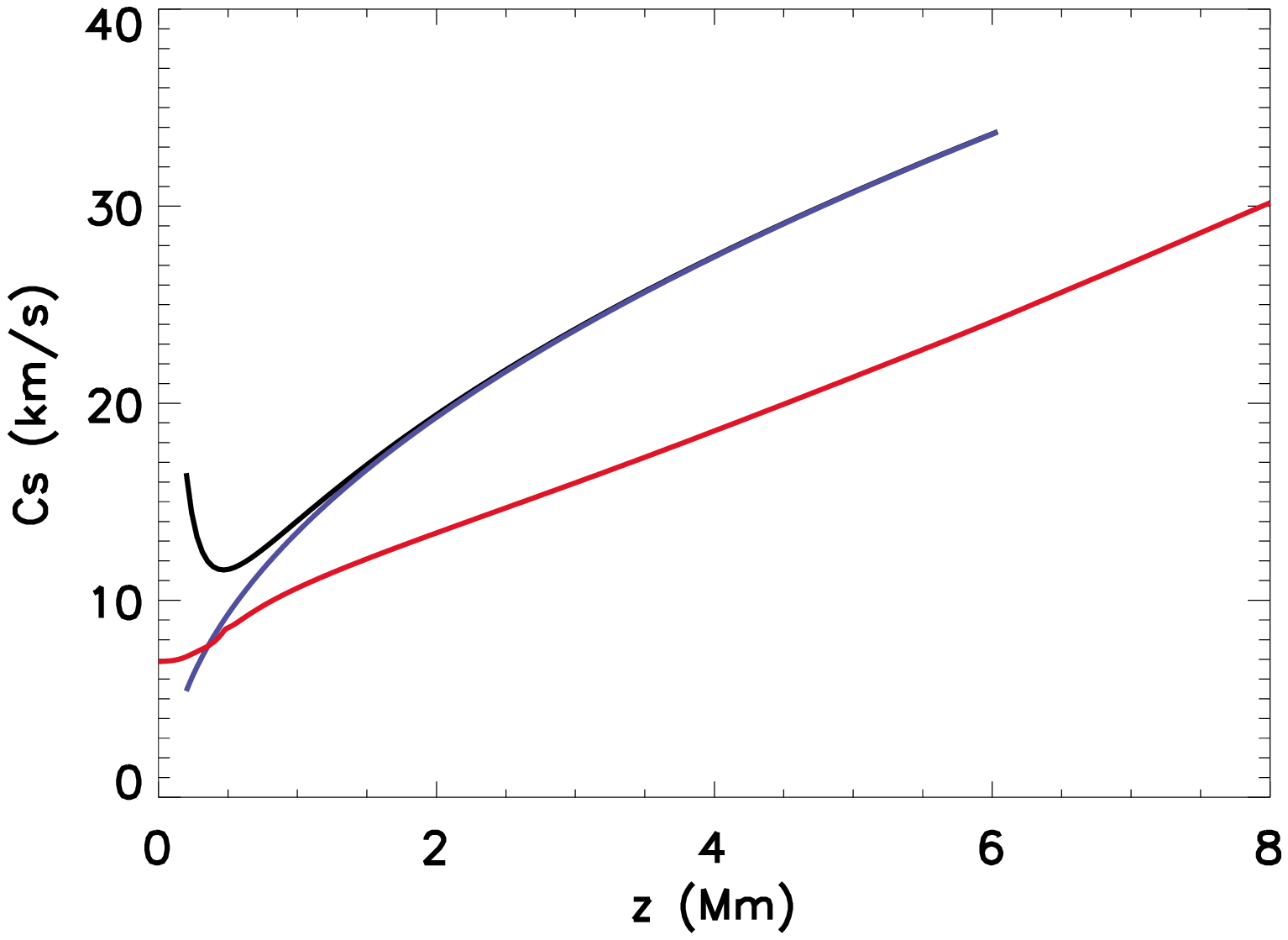}
\caption{ The vertical profiles of the pressure, density and acoustic velocity for the background model. Red plots are the profiles for the standard solar model S by Christensen-Dalsgaard (with $z = 0$ corresponding to the height of 400 km in model S). Blue plots are for the polytrope along the flux-tube axis, and the black plots are for the enhanced polytrope in quiet the Sun.}
\label{profiles}
\end{figure}

 We use a slightly modified version of the \textit{SLiM} (\textit{Semi spectral
 Linear MHD}) code described in \cite{Cameron07} to propagate linear waves through an enhanced polytropic
 atmosphere.
The enhanced polytropic atmosphere we use was constructed by following  \cite{Cally97}. 
Specifically we follow those authors in considering a 4820~G purely vertical flux tube 
along the axis of which the pressure, density, and temperature vary as a polytrope. 
The polytropic index is set as 1.5, the density at a depth of 1 Mm is set to $10^{-5}$ g cm$^{-3}$.
The sound speed is set to be equal to the Alfven speed at depth of 400 km. The external 
atmosphere is the enhanced polytrope obtained by requiring that the total (gas and magnetic) 
pressure be horizontal uniform, and by that the density is that of the original polytrope. 
Since the magnetic pressure is independent of height this 
means that away from the flux tube the pressure is increased by a comstant which is 
independent of height and thus does not affect the vertical hydrostatic force balance. We emphasise that this is the same formulation as was used by \cite{Cally97} and is a 
reasonable approximation for qualitatively studying wave propagation in the upper part 
of the convection zone.
 Figure~\ref{profiles} show a comparison between the vertical profiles of the sound speed, pressure, and acoustic velocity for the standard solar model S by Christensen-Dalsgaard and those on the flux-tube axis. We note that with our boundary conditions waves are strongly reflected from the upper boundary 
which restricts us to considering waves below the acoustic cut-off frequency.
As in \cite{Cally97}, we restrict ourselves to purely vertical flux tubes. In our case this is partly motivated by
the fact that we are considering only small tubes (not sunspots) which are expected to
be almost vertical until they expand into the overlying atmosphere.
We thus are considering relatively small flux tubes of different sizes and are including 
the full three-dimensional geometry.

Our code solves the linearized, ideal MHD equations in three dimensions. We write the magnetic, velocity, 
pressure, and density perturbation in terms of the displacement vector [$\bxi$]. The background is assumed 
to be time independent. The primed quantities in the equations are the Eulerian perturbation, whereas the 
background variables are subscripted with 0. The gravitational acceleration  [$\bg$]  is constant, and there 
is no background steady flow. 

The equation governing the displacement vector is:

\begin{equation}
\rho_{0}\frac{\partial^{2}
\bxi}{\partial t^{2}}  = -\grad p'+ \rho'\bg+ \frac{1}{4\pi} (\jj'\times\bb_{0}+\jj_{0}\times\bb')
\end{equation}

The perturbed quatities:  density [$\rho$], pressure [$p$], magnetic
field [$\bb$]. Electric current [$\jj$] are written as functions of [$\bxi$]  as follow:

\begin{equation}
\rho'=-\grad \cdot(\rho_{0}\bxi) ,
\end{equation}

\begin{equation}
\label{pip}
p'=c_{0}^2 (\rho'+\bxi\cdot\grad\rho_{0})-\bxi\cdot\grad p_{0} ,
\end{equation}

\begin{equation}
\label{Bip}
\bb'=
\curl(\bxi\times\bb_{0}) ,
\end{equation}

\begin{equation}
\label{Jip}
\jj'=
\curl\bb' ,
\end{equation}

A flux tube that is almost evacuated at the surface and that has a
radial profile is superposed on the background atmosphere. The
magnetic flux tubes have a top-hat profile given by $ B(r) = B_{0} \exp(- r^4/ r_0^4) $ where $r_{0}$ 
is the tubes radius, and $B_{0}$ = 4820 G  \citep{Cally97}. We use a local Cartesian geometry defined by the 
horizontal coordinates $x$ and $y$. The vertical coordinate is $z$, and the simulation covers the height range from 0.2~Mm to 6~Mm below the solar surface. Note that the convention adopted here is that $z$ incresaes with depth. 
The horizontal extent of the domain is $x$  $\in$ [-20,20] Mm and  $y$  $\in$ [-10,10] Mm. 
The axis of the vertical flux tube passes through the point $x$ = -7 Mm, $y$ = 0.

At $t_0 = 0$, an $f$-mode wave packet with a Gaussian profile in Fourier
space is situated at $x_0 $ = -20 Mm and it propagates from left to right
in the $x$-direction. 

Specifically the initial condition is constructed
from individual $f$-modes each of which has the usual (wavenumber-dependent) 
exponential dependence on height. The phase of the each eigenmode is 
set so that the waves are all in phase at $x = -20$ Mm. 
The distance from the initial
condition and the sunspot was chosen as the result of a compromise: On the one hand
it is desirable for the initial wavepacket to be close to the sunspot (so that the
wave attenuation before the wave reaches the sunspot is minimized), on the other hand
there is a relationship between the wavepacket and cross covariance from random sources
\citep{Cameron08,Gizon10} which only applies outside the ``near field'' 
of the initial condition.
The frequency of each mode is $\sqrt{g k}$ where $k$ independently
of the details of the background quiet-Sun model. The vertical component of the
velocity, $\partial_t\xi_z$, at the upper surface is our main diagnostic, as it is closely 
related to the observable Doppler velocity at disk centre. The amplitude of the 
different modes in the initial condition depends on their eigenfrequency. The amplitudes
of the surface velocity is a Gaussian with a central frequency and the half width of the wave packet are 3 mHz and 1.18 mHz respectively. 

The top boundary also corresponds to that of \cite{Cally97} and has
also been previously described in \cite{Cameron08}. It corresponds to a free surface in the non-magnetic
regions and hence is effective in reflecting waves. This means that the effective acoustic cut-off of the
model is infinite. Our initial condition has very little energy ($\approx$ 0.06\% above 5.3 mHz).

We have also performed a simulation without the flux tube being present, that is
with the enhanced polytrope being used everywhere and with $B_{0}=0$. This simulation
acts as a reference and allows us to construct the scattered wave field as the difference
between the simulations with and without the flux tube.

\begin{figure}    
\centering
\includegraphics[width=0.36\textwidth,clip=]{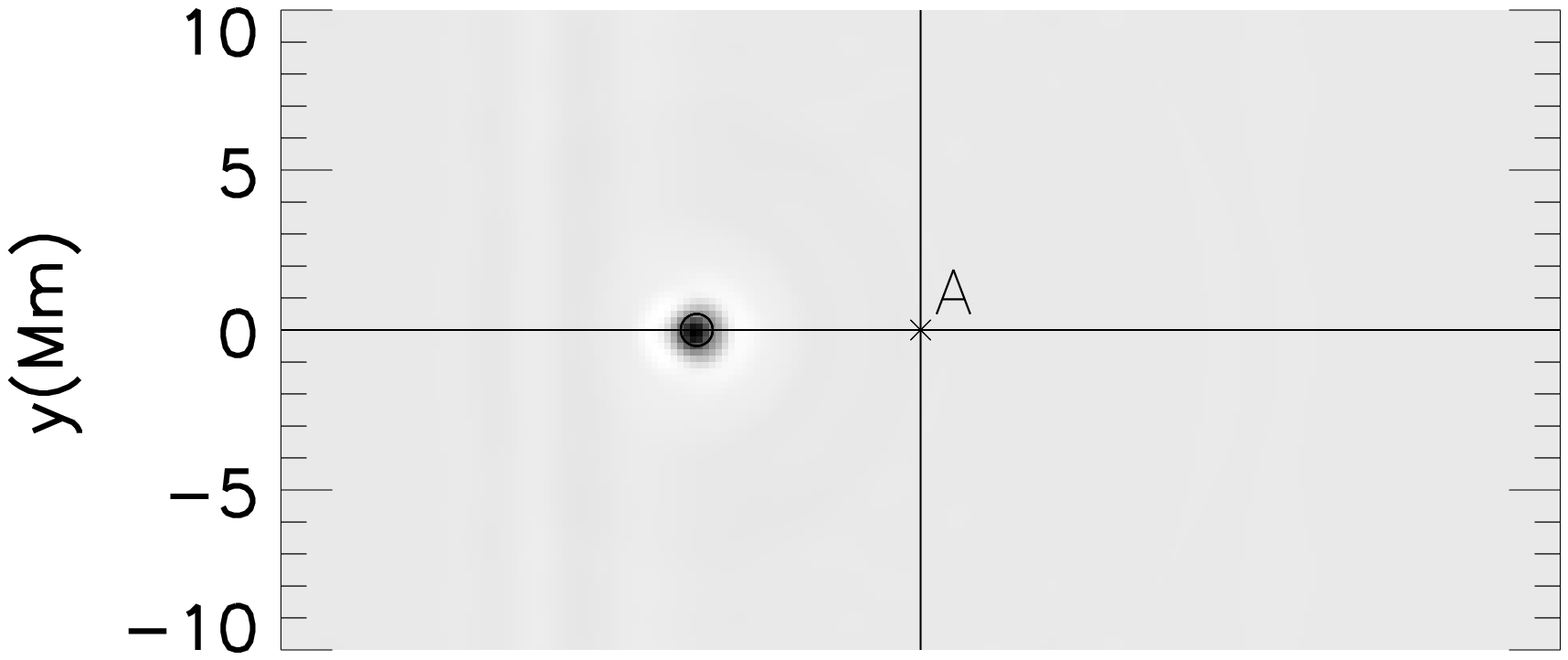}
\vspace{0.02\textwidth}
\includegraphics[width=0.30\textwidth,clip=]{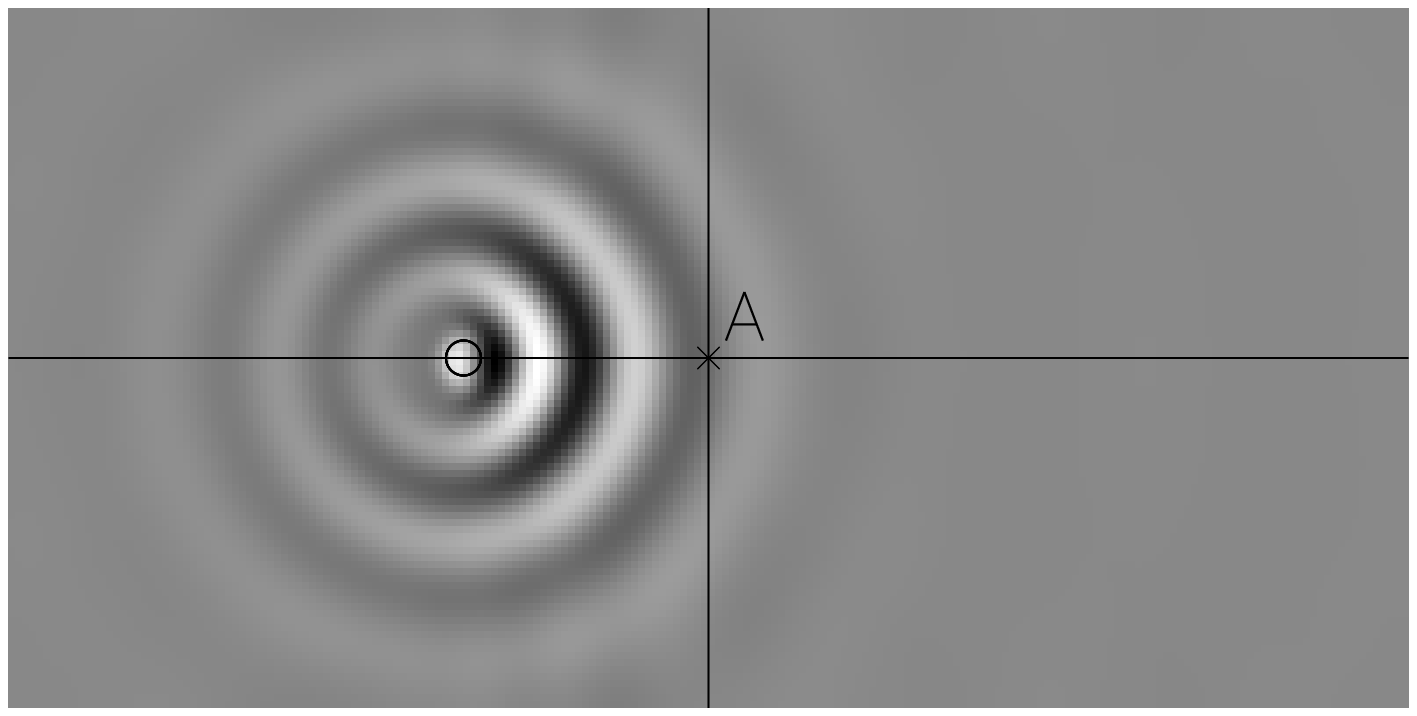}
\includegraphics[width=0.30\textwidth,clip=]{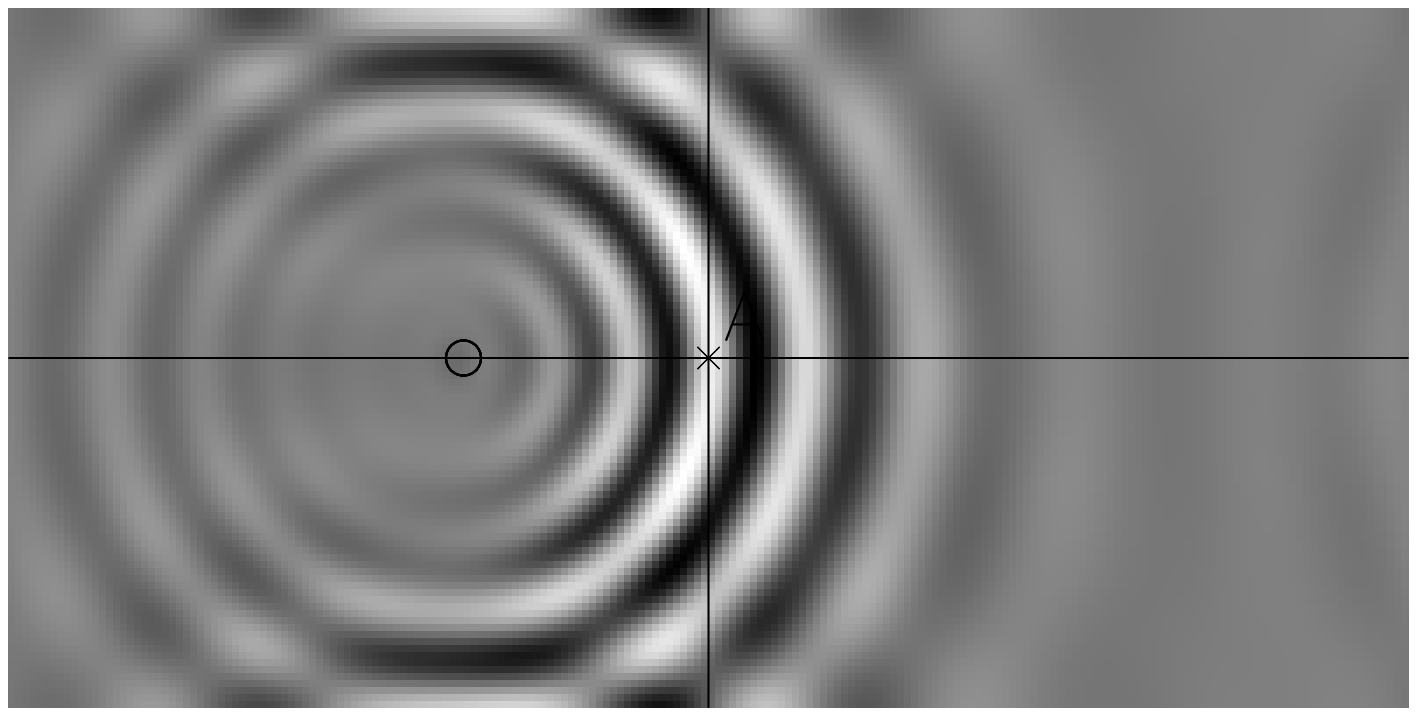}
\includegraphics[width=0.35\textwidth,clip=]{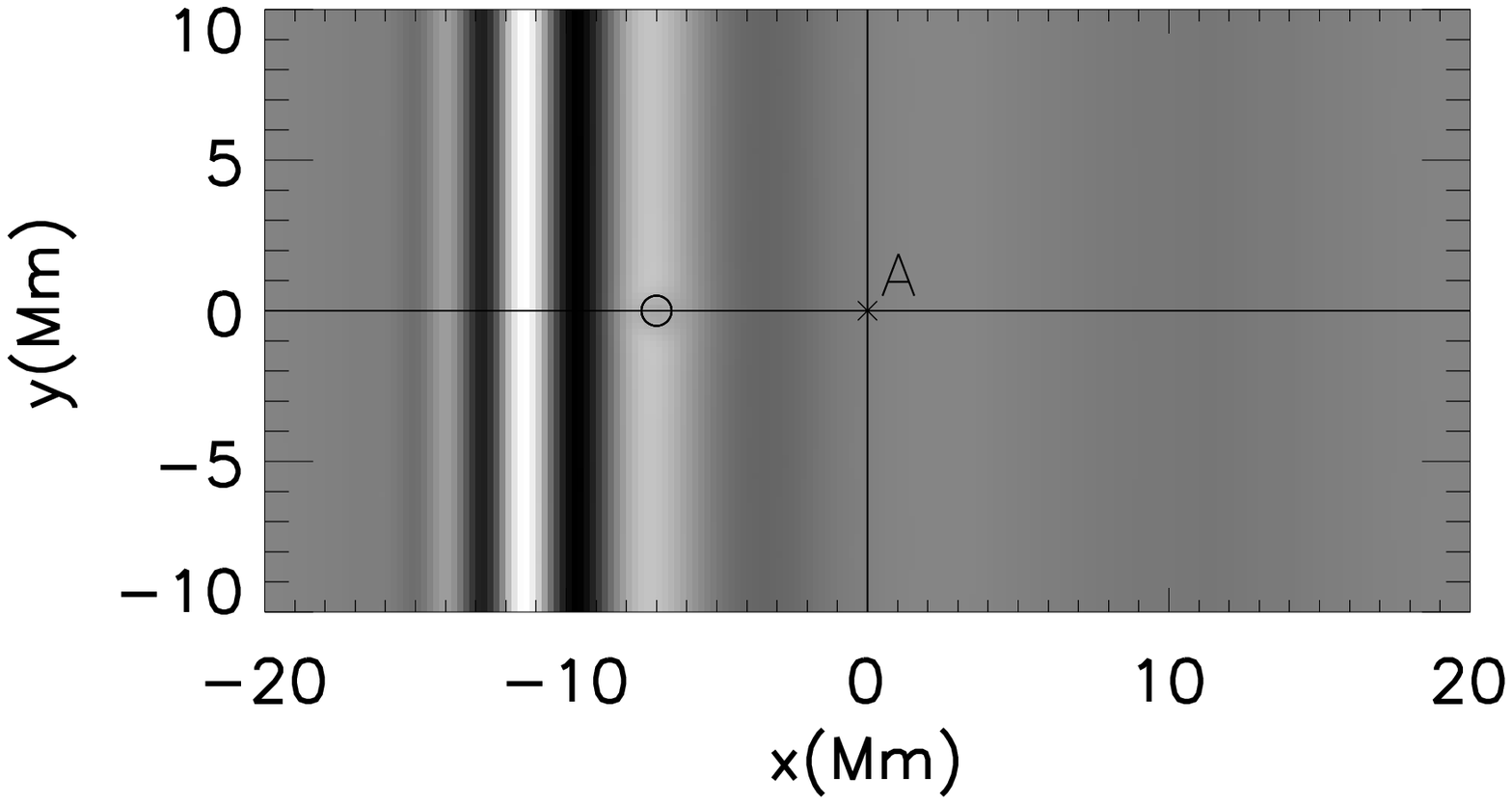}
\vspace{0.05\textwidth}
\includegraphics[width=0.30\textwidth,clip=]{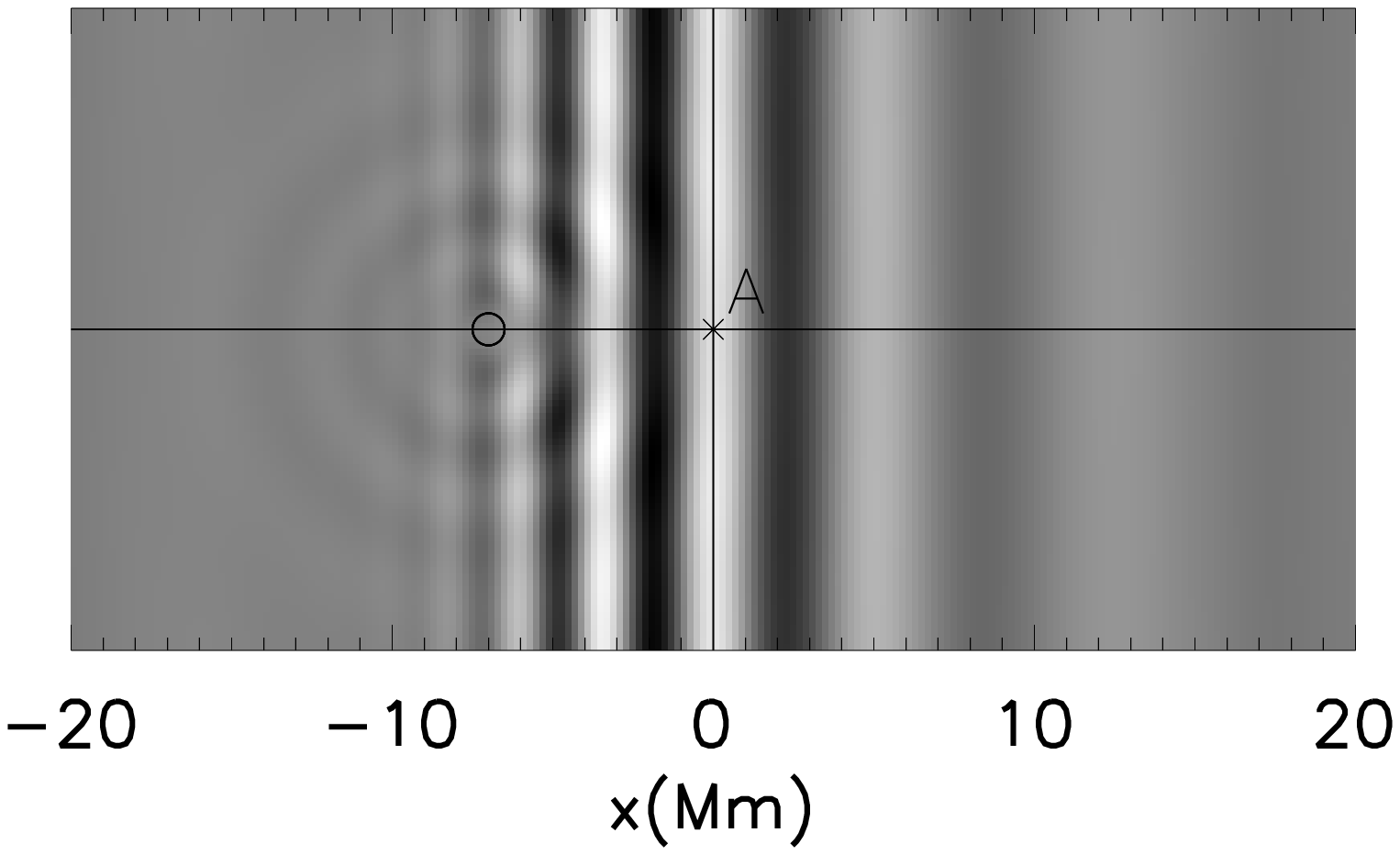}
\includegraphics[width=0.30\textwidth,clip=]{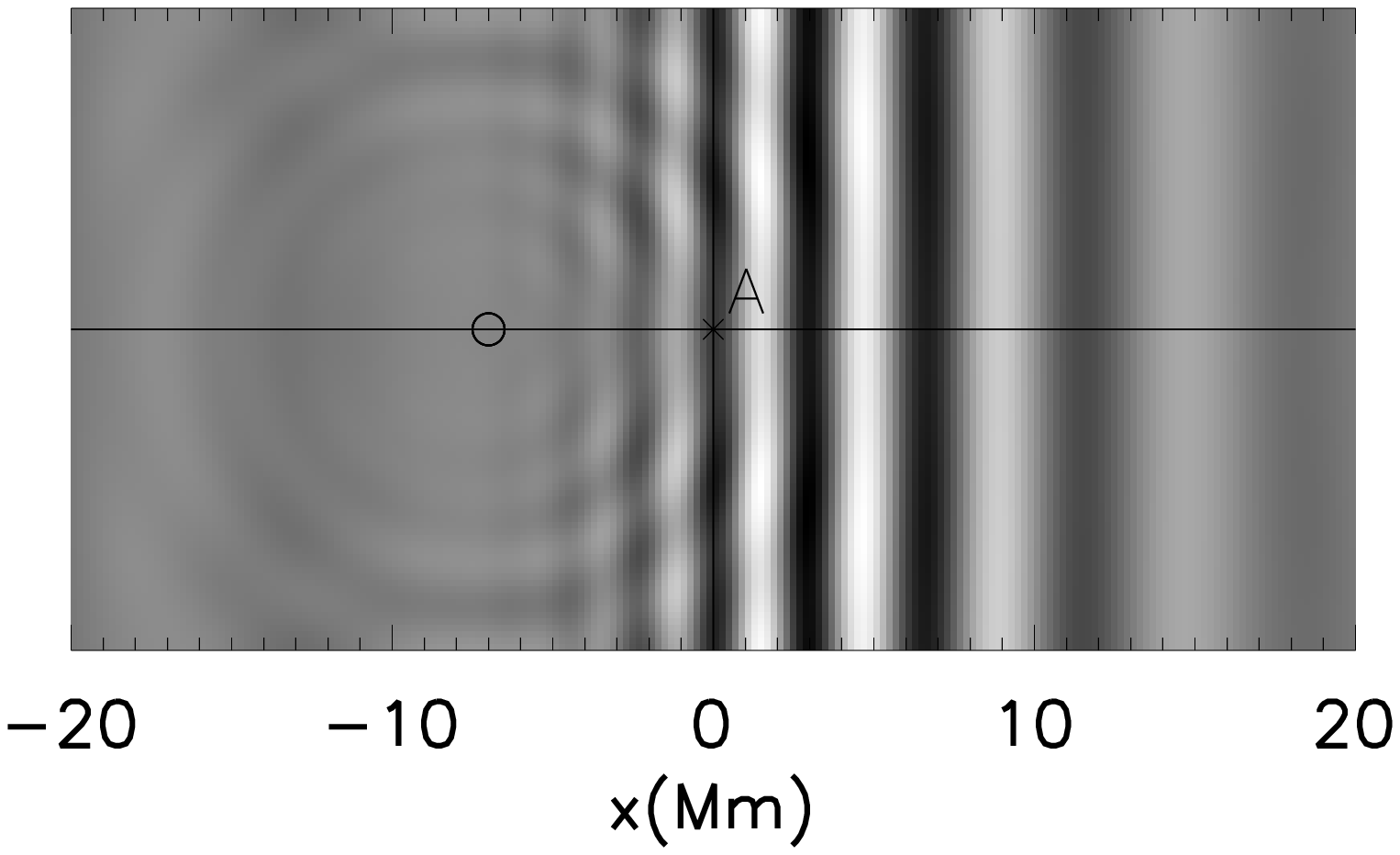}
\includegraphics[width=0.34\textwidth,clip=]{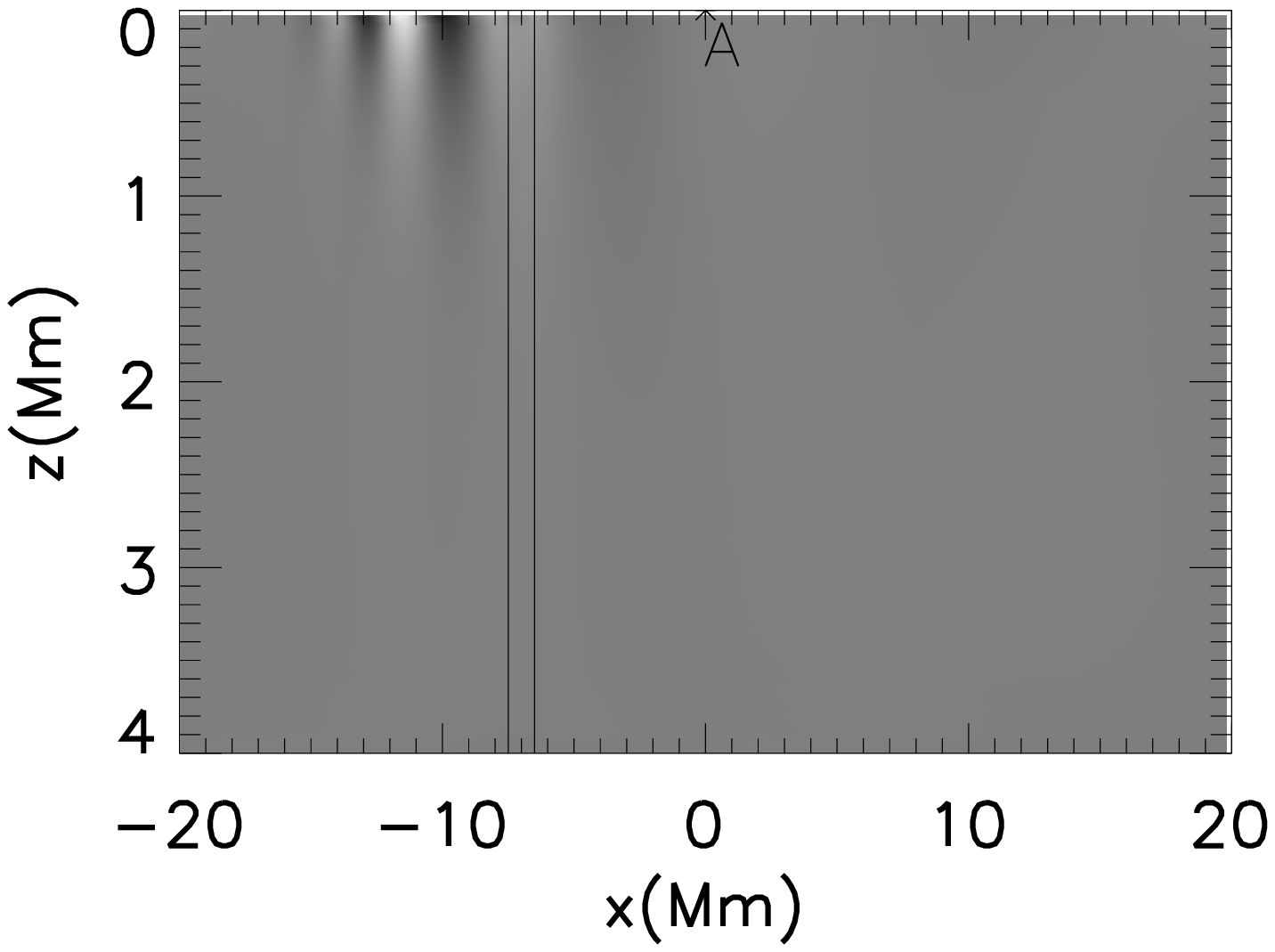}
\includegraphics[width=0.31\textwidth,clip=]{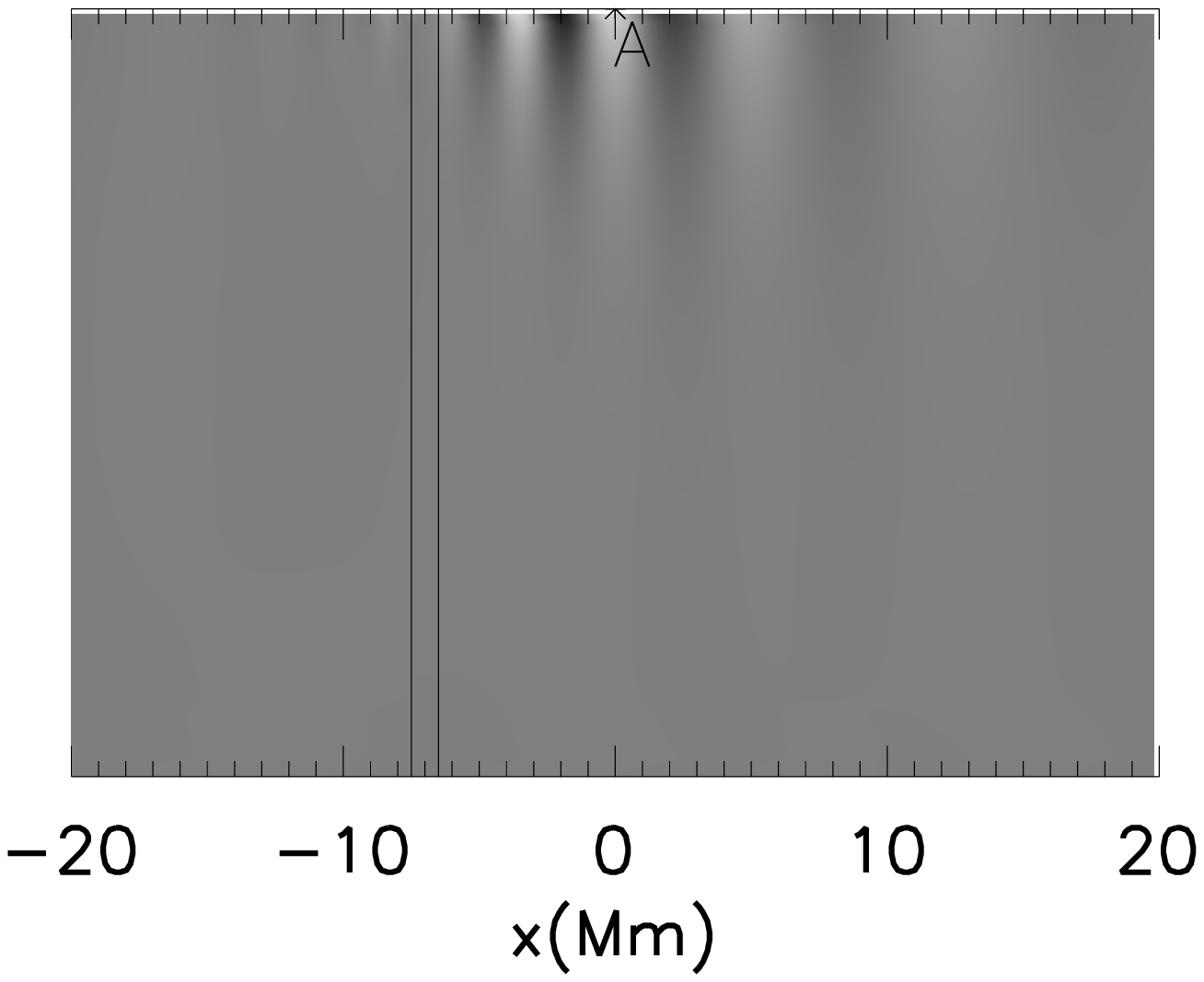}
\includegraphics[width=0.31\textwidth,clip=]{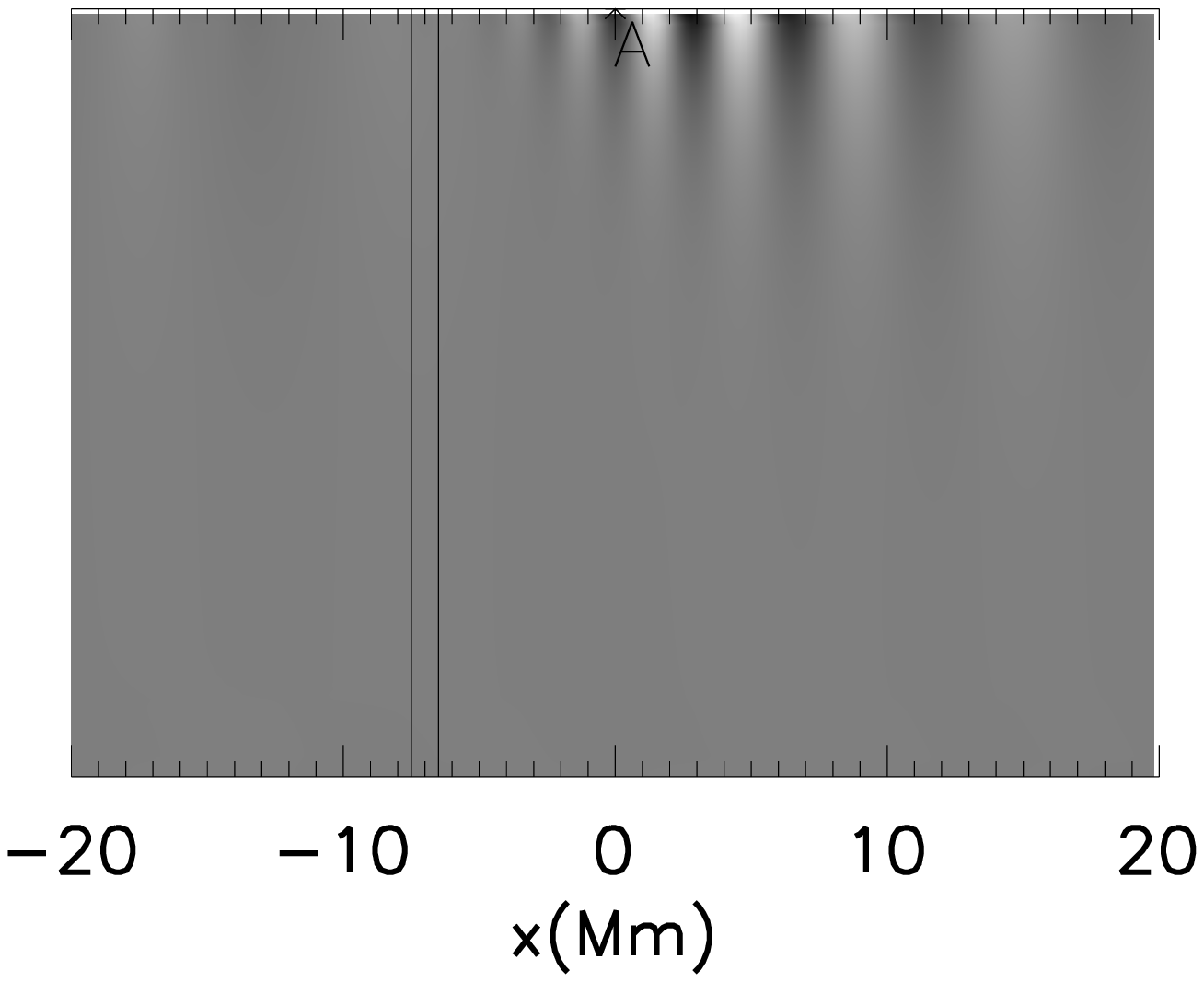}

\caption{ Wave propagation at three times (1600, 3000, and 4000 seconds
  after the start of the simulation), in the presence of a 500 km
  magnetic flux tube. The top panel shows the scattered component of
  the wave field near the ``solar surface'', the middle panel shows
  the full wave field whilst the lower panel is a vertical cut. The
  location and size of the tube is indicated by the circle at $x$ = -7 Mm, $y$ = 0. The point ``A'' is the location at which the
  resolution test analysis was performed (Section~\ref{S-Resolution test}). Notice that the simulation is
periodic in the horizontal directions, which can be seen in the scatterd wave field at later times.}
\label{hr500}
\end{figure}

An example of the wave interacting with a tube with a radius of 500 km is shown in Figure~\ref{hr500}.
For a tube of this size, the scattered wave is almost circular, with a significant amount of back scattering. The \textit{SLiM} code assumes that the box is periodic, which accounts for the wave fronts coming in from the side of the box after 4000 seconds. It is to be noted that the diameter of the tube is neither small nor large with respect to the wavelength of the incoming waves, so that various types of tube modes will be excited.

\section{Resolution Test}
\label{S-Resolution test}
\begin{figure}   
\centering
\includegraphics[width=1\textwidth]{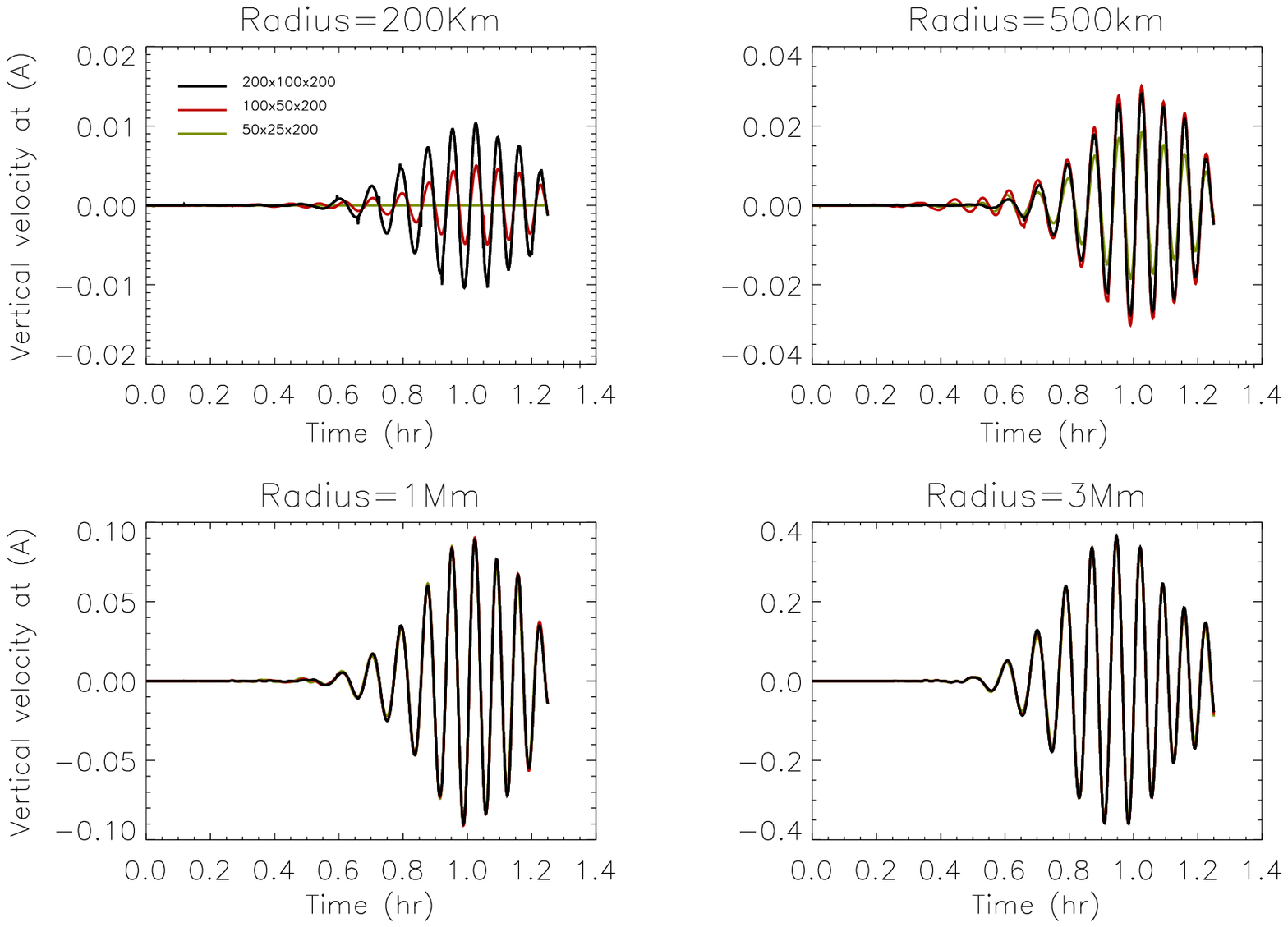}
\includegraphics[width=1\textwidth]{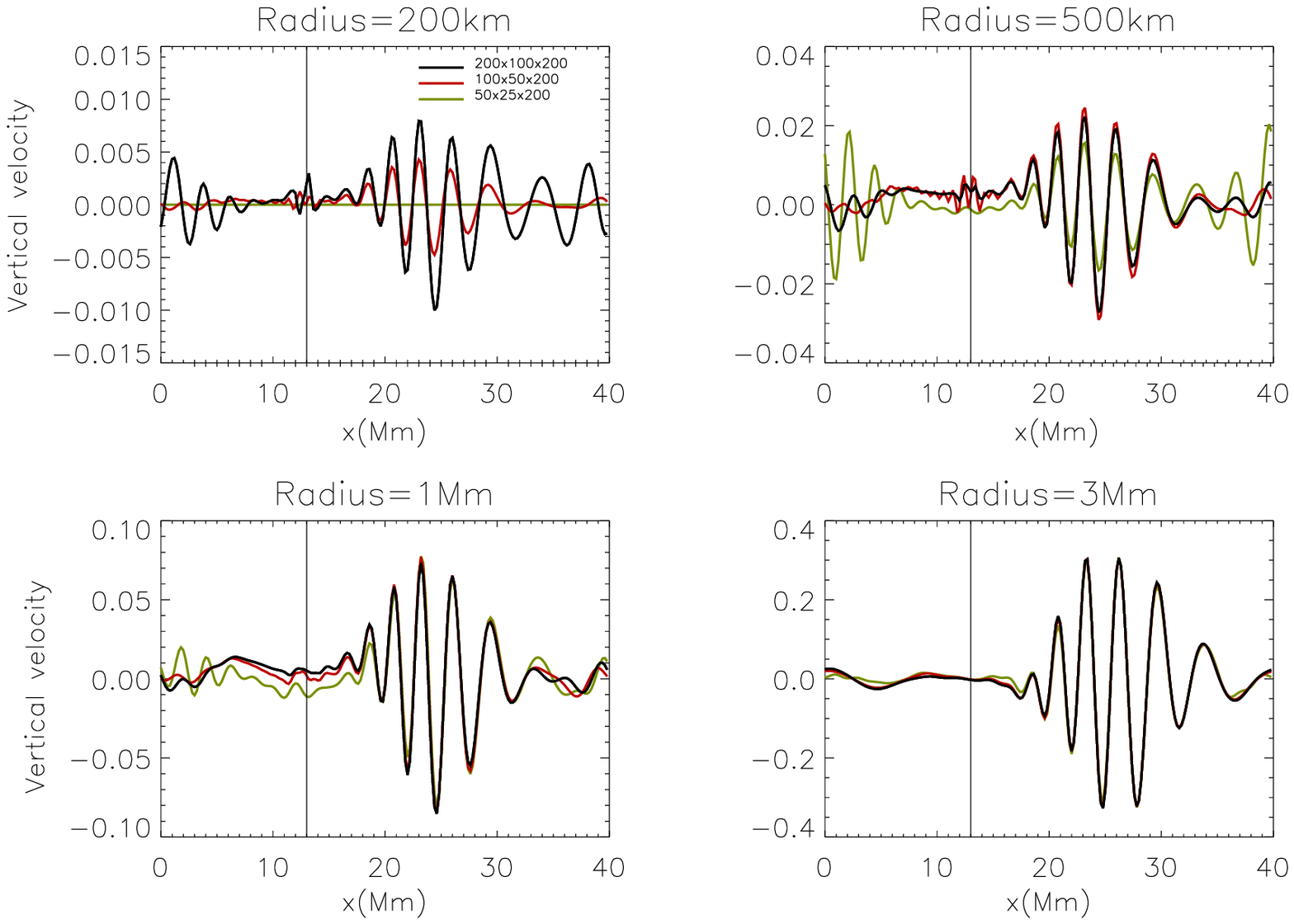}
\caption{\ Comparison of results at different resolutions for various
  sizes flux tubes. The top four panels show the scattered component of
  the wave at a fixed point (``A''in Figure~\ref{hr500}) as a function of
  time. The lower four panels show the scattered wave component at
  different locations along the $x$-axis at fixed time (corresponding to
  the middle panels in Figure~\ref{hr500} )}
\label{restx}
\end{figure}

Our simulation box has a length of 40 Mm in the direction in which the
wave is propagating. The radii [$r_o$] of the flux tubes we are
interested in vary from 200 km to 3 Mm. Furthermore, since we have used a
top-hat profile [$ B(r) = B_{0} \exp(- r^4/ r_0^4) $] the ``sidewalls''
of the tube are even thinner. This raises the question of how many
Fourier modes are required in the simulation to resolve the
flux tubes. In the vertical direction we have found that 200 grid
points are sufficient for all the flux tubes, probably because the flux tubes have no special structure in the $z$-direction. The tubes radii we have considered are 200 km, 500 km, 1 Mm, and 3 Mm.
We have considered three different choices for the number of horizontal Fourier modes: 50$\times$25, 100$\times$50, 200$\times$100. The data of the scattered wave field at the point ``A'' is shown as a function of time in the top four panels of Figure~\ref{restx}. The three different resolutions are shown using different colors.
It is apparent that the scattered wave at ``A'' is resolved for $r_0$
equal to 1 Mm and 3 Mm for all three choices of the number of Fourier
modes. The degree of convergence in the $r_0=500$ km case
is reasonable, especially for the two higher resolutions.
The case where $r_0 = 200$ km is clearly a long way from convergence for
at least the two lower resolutions. At the lowest resolution, the flux
tube is significantly smaller than $40$ Mm$/50$ modes $=800$ km, for this reason the simulated waves apparently do not even see the tube and the scattered wave is absent. With 100 modes, the simulation appears more reasonable but the amplitude is only about 1/2 of that of the simulation with 200 modes. The results using 200 modes should be much closer to the converged value: The relative resolution is similar to the case where we used 100 modes for the 500 km tube.
The four lower panels of Figure~\ref{restx} show the scattered wave along the line $y=0$ as a function of $x$ at a fixed time. The convergence properties are the same as we found for fixed position ``A''.
To quantify the errors we introduce a measure for the difference between the simulations. Again concentrating on the point ``A'', we compare the time series of the vertical velocity from either the low or medium-resolution simulations [$v_z$] with the high resolution simulation [$v_z^h$]:

\begin{equation}
\epsilon=\frac{\sum_{t}[v_{z}(t)-v_{z}^{h}(t)]^{2}}{\sum_{t}v_{z}^{h}(t)^{2}}   ,
\end{equation}

We can also construct a measure of the error based upon the cuts at fixed time (\textit{i.e.} the lower four panels of Figure~\ref{restx}). The error in this case is 
taken to be

\begin{equation}
\epsilon=\frac{\sum_{x}[v_{z}(x)-v_{z}^{h}(x)]^{2}}{\sum_{x}v_{z}^{h}(x)^{2}}   ,
\end{equation}

\begin{figure}   
\centering
\includegraphics[width=1\textwidth]{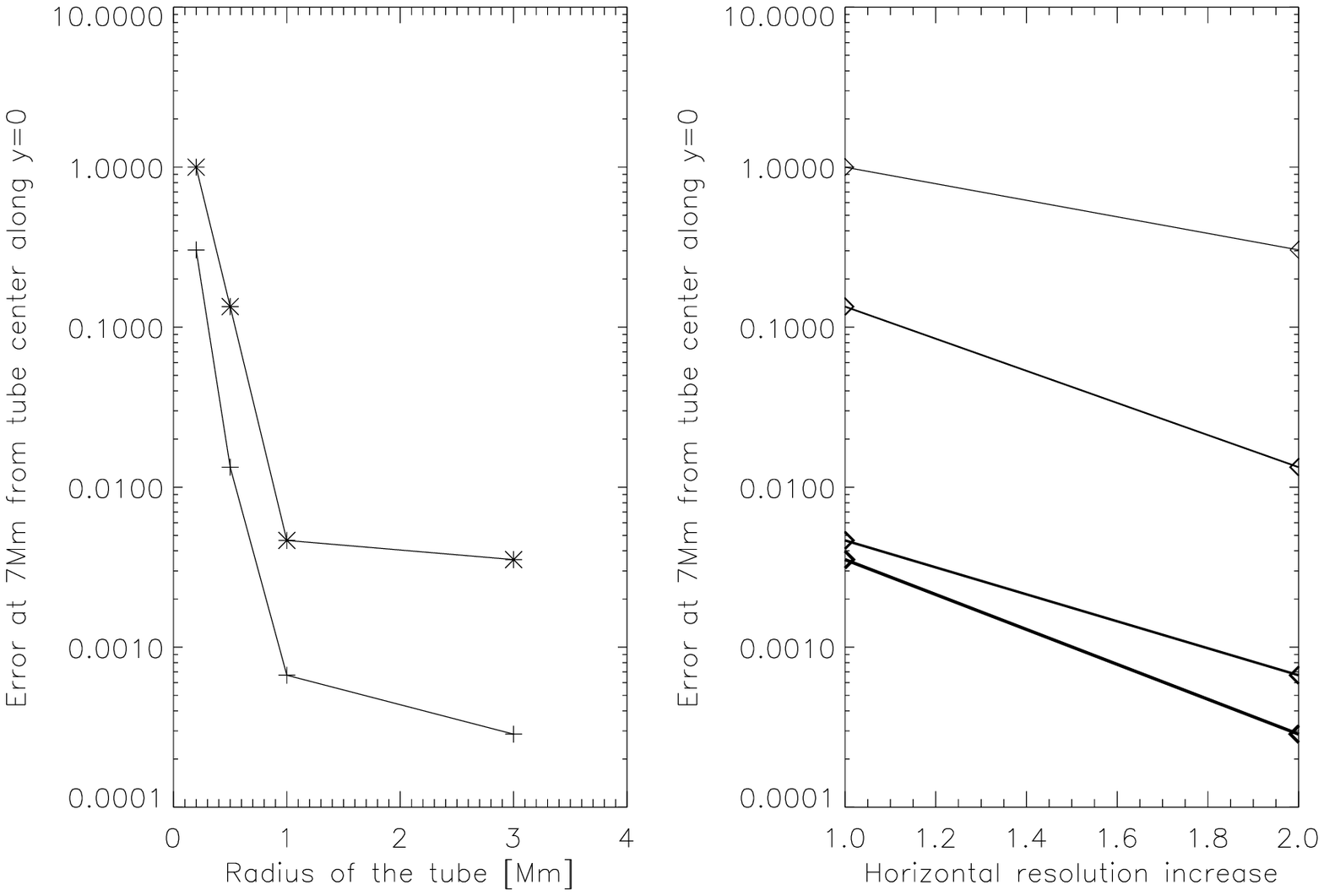}
\includegraphics[width=1\textwidth]{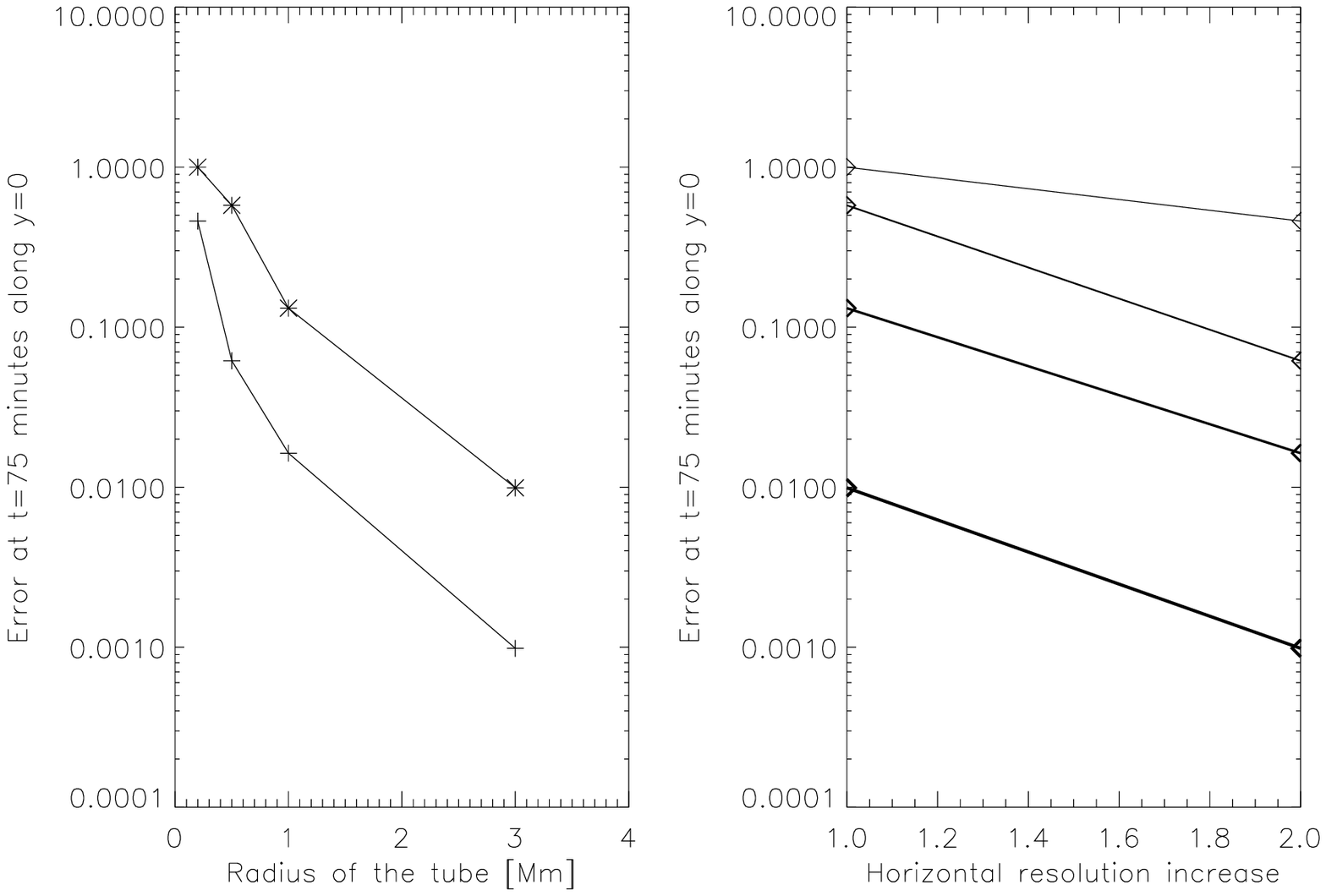}
\caption{\ The error function for scattered wave field of vertical velocity, top panel at 7 Mm from tube center along $y$ = 0, bottom panel at time = 75 minutes along  $y$ = 0.  Curve with stars at left correspond to the difference low \textit{vs} high resolutions, curve with cross is for the difference medium \textit{vs} high resolutions. The right curves shows the variation of the error for the different radius when the horizontal resolution increases.}
\label{restx2}
\end{figure}

Figure~\ref{restx2} shows these measures of the error in various ways. In the left panels we see that the  error decreases rapidly as the tube increases in radius. The panels on the right show that increasing the resolution by a factor of two decreases this measure of the error by a factor of about ten. This would suggest that the error associated with the highest resolution (200 modes) when applied to the smallest flux tube (200 km radius) should be approximately 3\% (\textit{i.e.} a factor of ten better than the simulation with 100 modes which has a 30 \% error).

\section{Scattering of Tubes }
\label{S-Scattering of tubes}
\begin{figure}    
\centering
\includegraphics[width=0.51\textwidth,clip=]{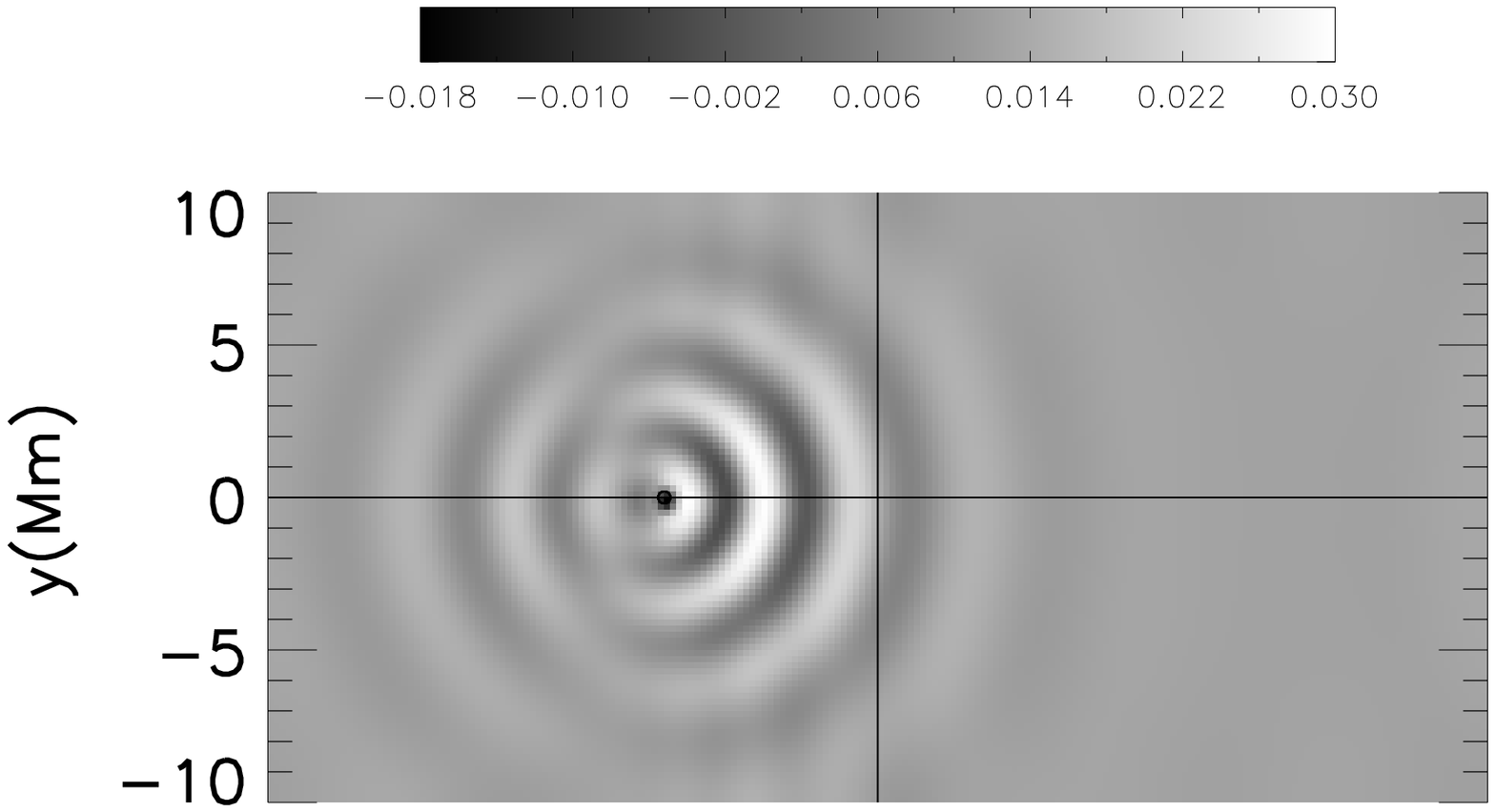}
\includegraphics[width=0.44\textwidth,clip=]{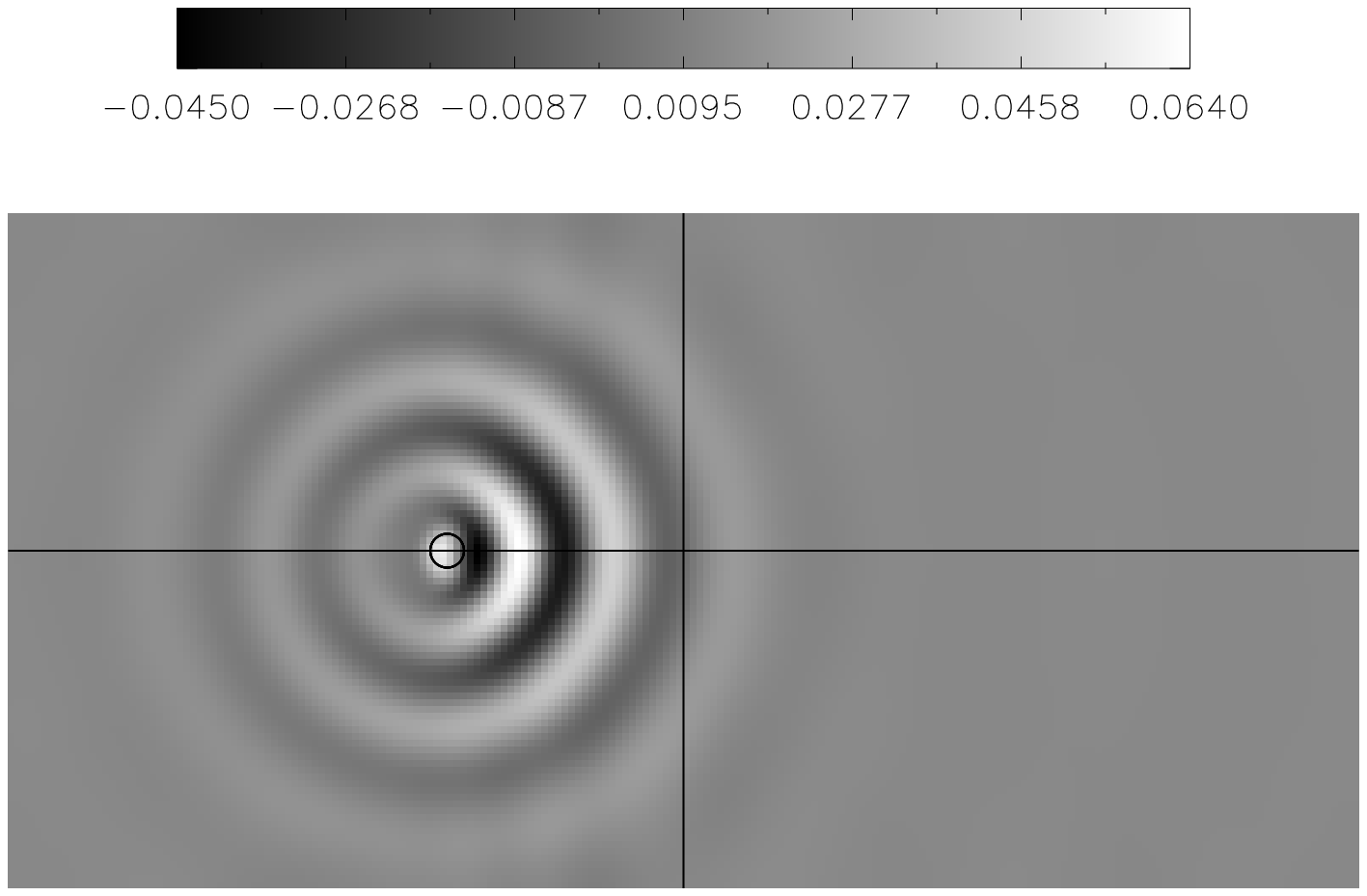}
\includegraphics[width=0.51\textwidth,clip=]{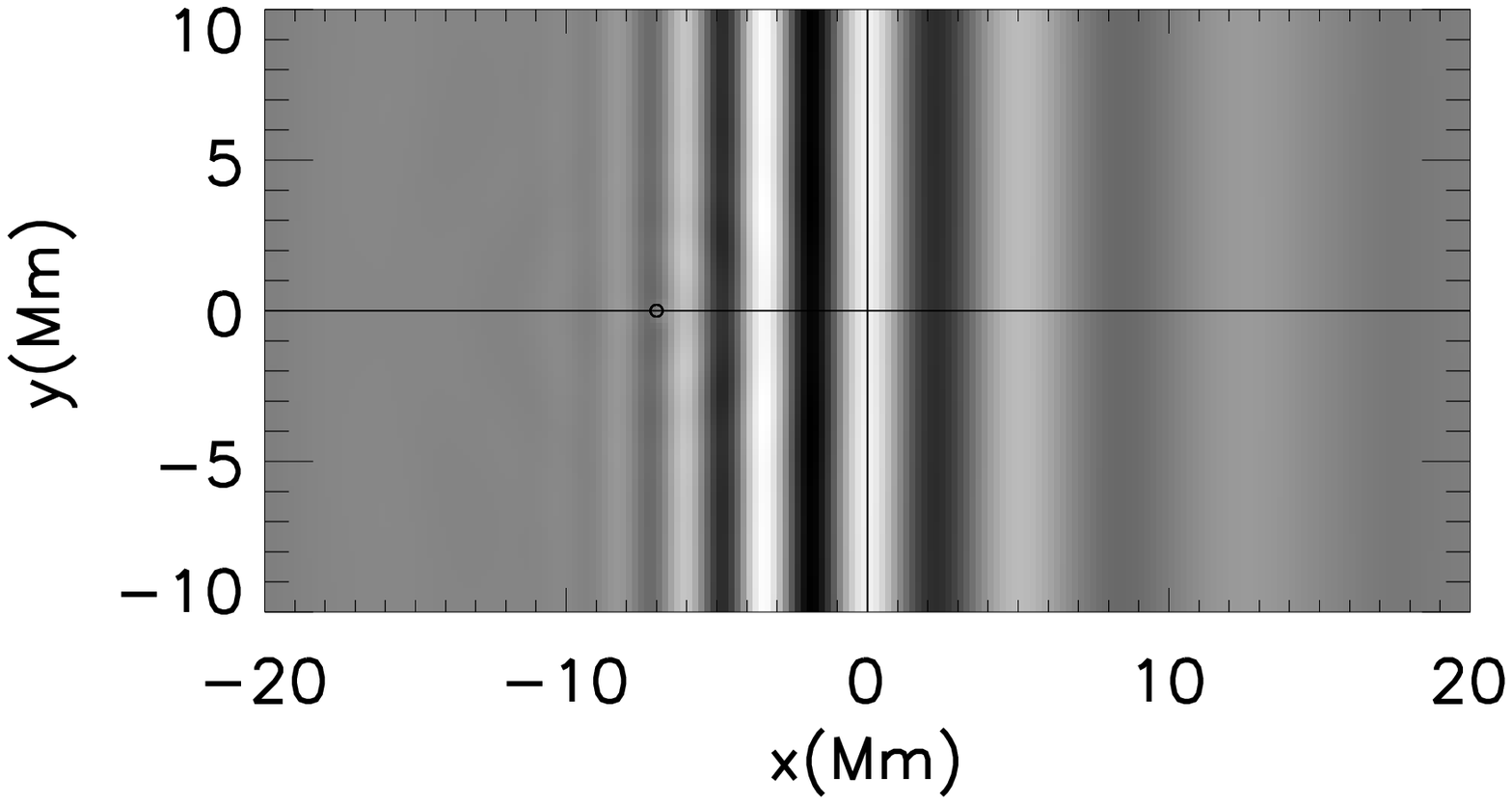}
\includegraphics[width=0.445\textwidth,clip=]{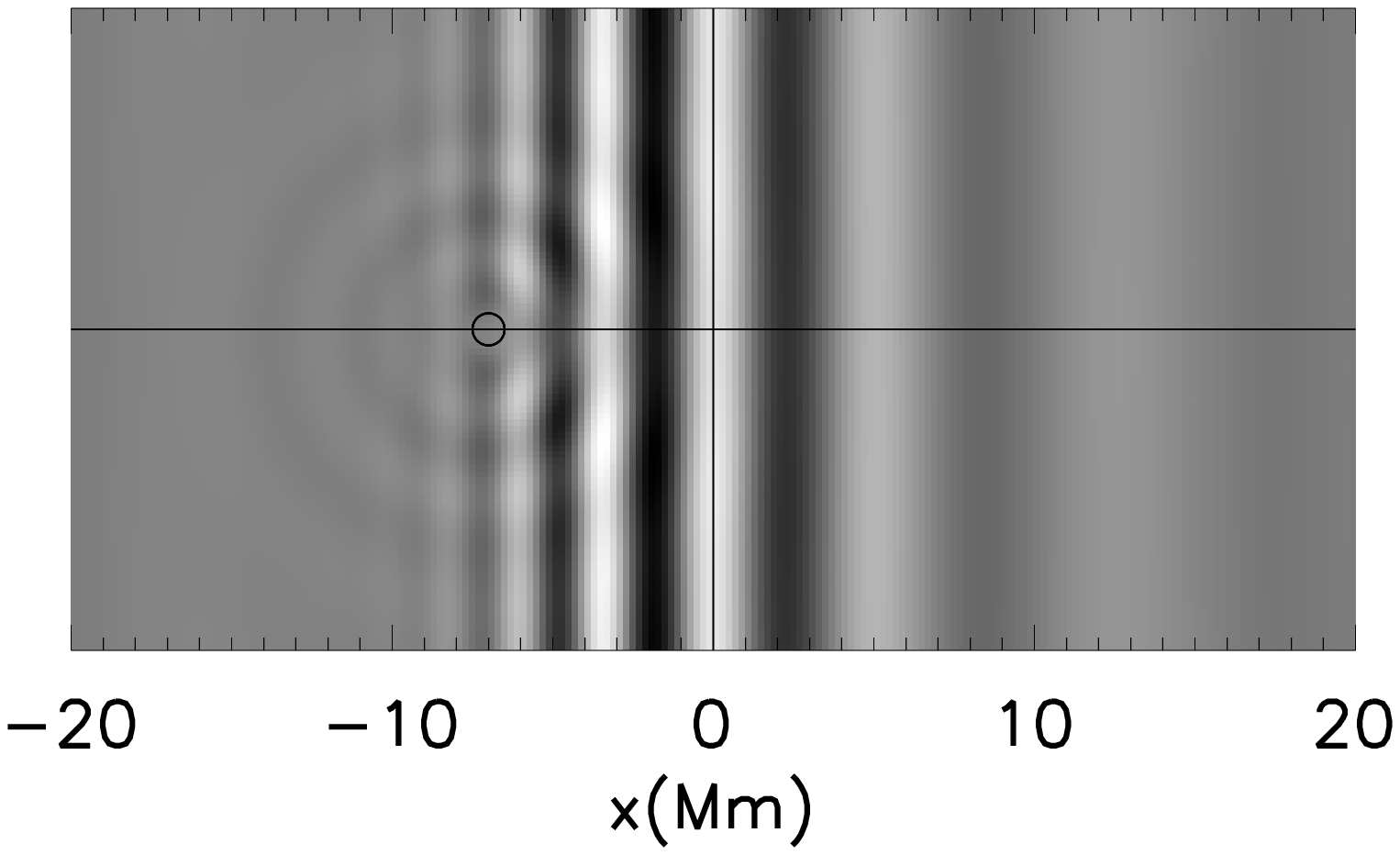}

\vspace{0.05\textwidth}
\includegraphics[width=0.51\textwidth,clip=]{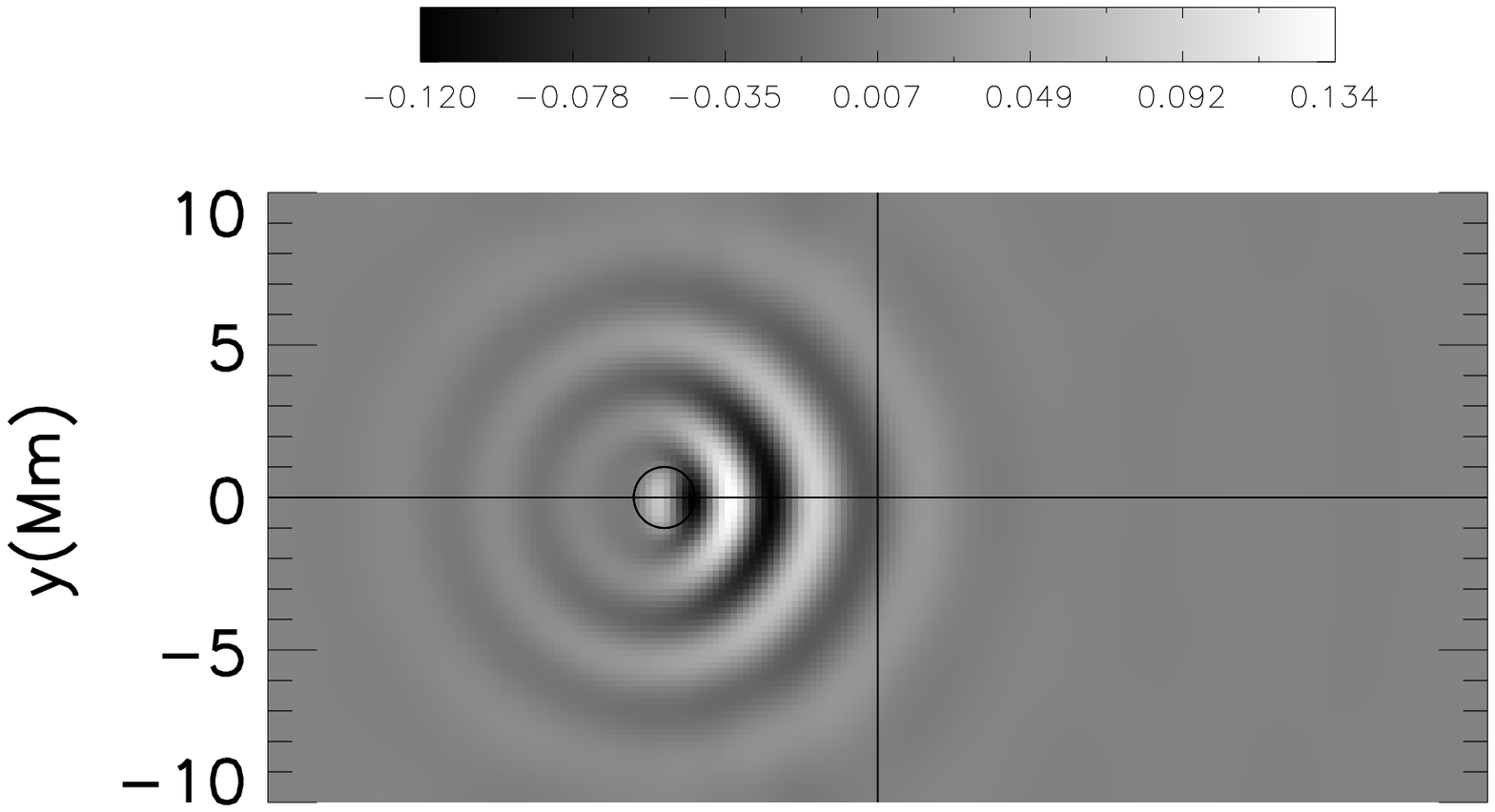}
\includegraphics[width=0.44\textwidth,clip=]{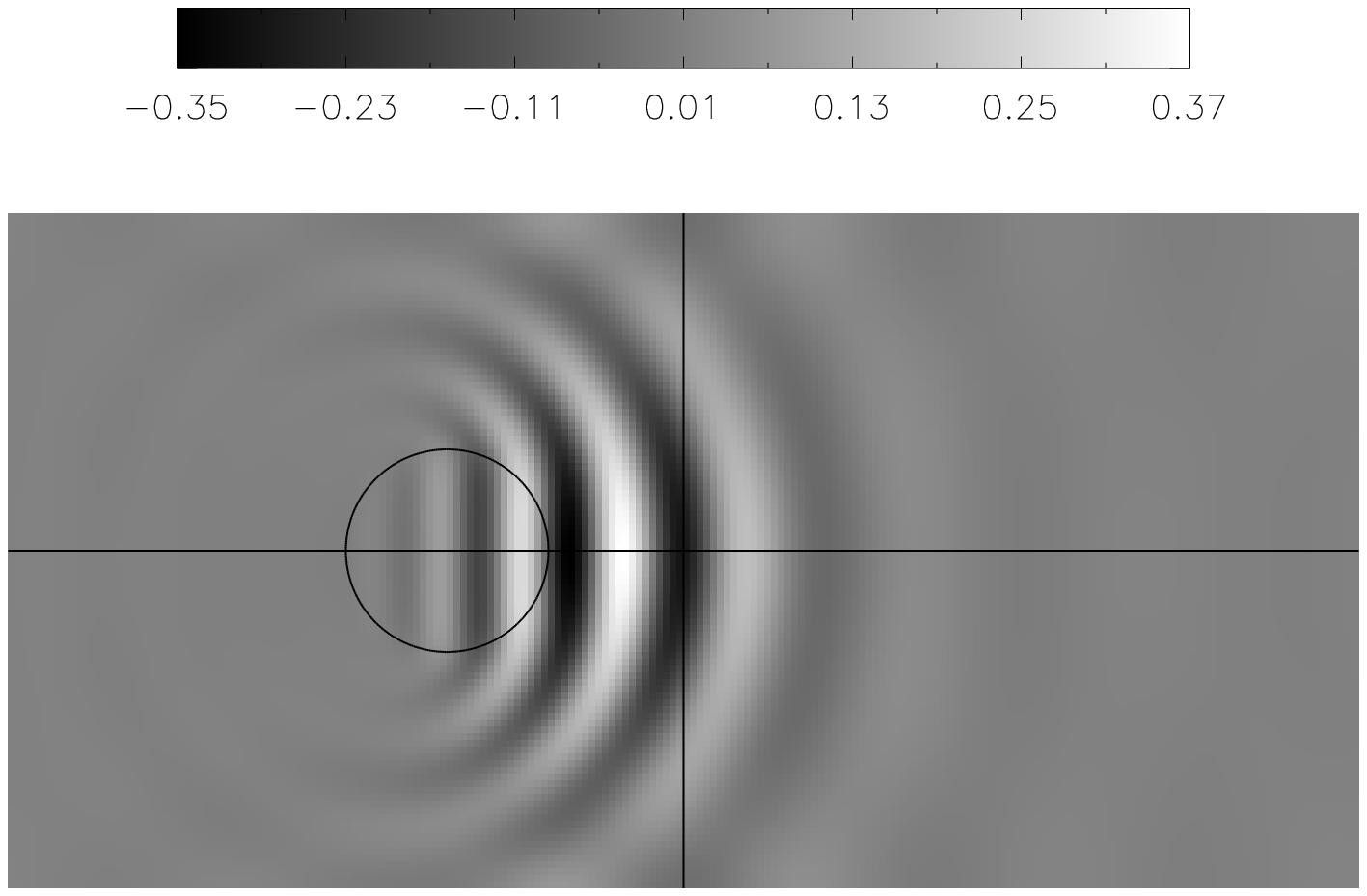}

\includegraphics[width=0.51\textwidth,clip=]{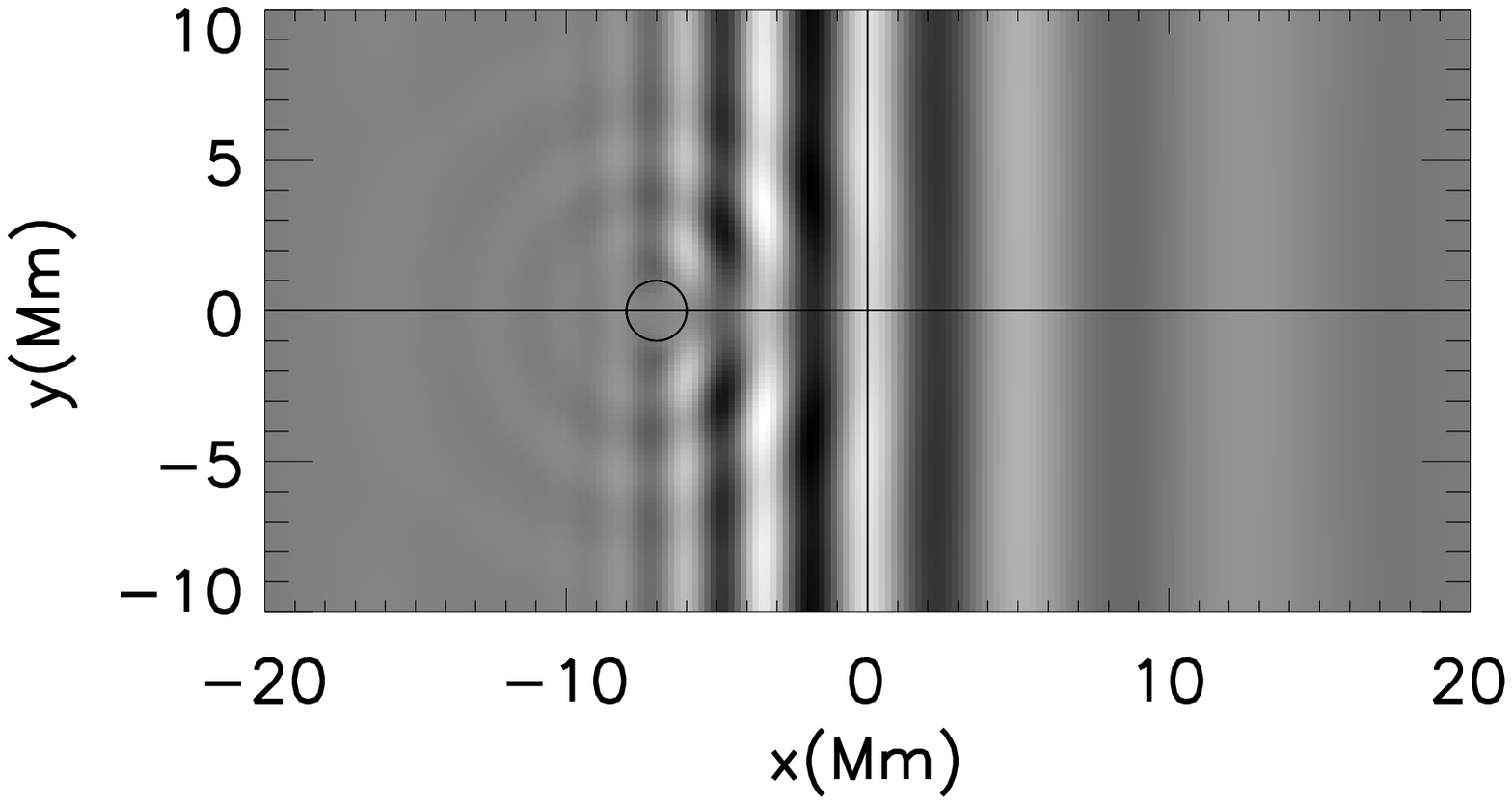}
\includegraphics[width=0.445\textwidth,clip=]{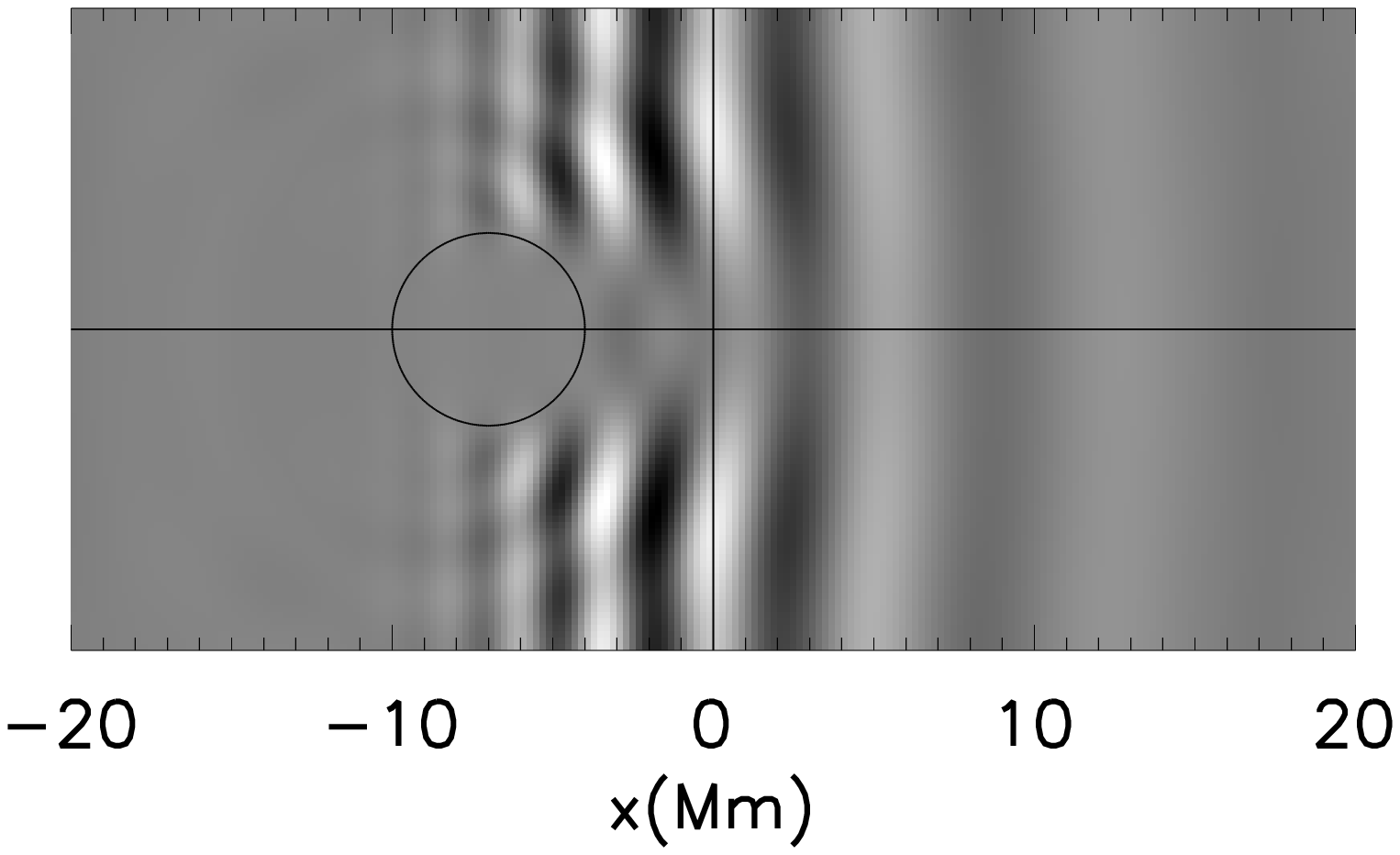}
\caption{ Wave propagation through tubes of different sizes. The
  images are from snapshots taken 3000 seconds after the start of the
  simulation. The upper four panels show vertical velocity scattered wave
  field (at the top) and the full wave field (at the bottom) for tubes
  with radii 200 km and 500 km, the lower four panels are for tubes with
  radii 1 Mm and 3 Mm. The circles indicate the sizes of these tubes. The simulation is
periodic in the horizontal directions which can be seen in the scattered wave field.}
\label{4rxy}
\end{figure}

\begin{figure}   
\centering
\includegraphics[width=0.32\textwidth]{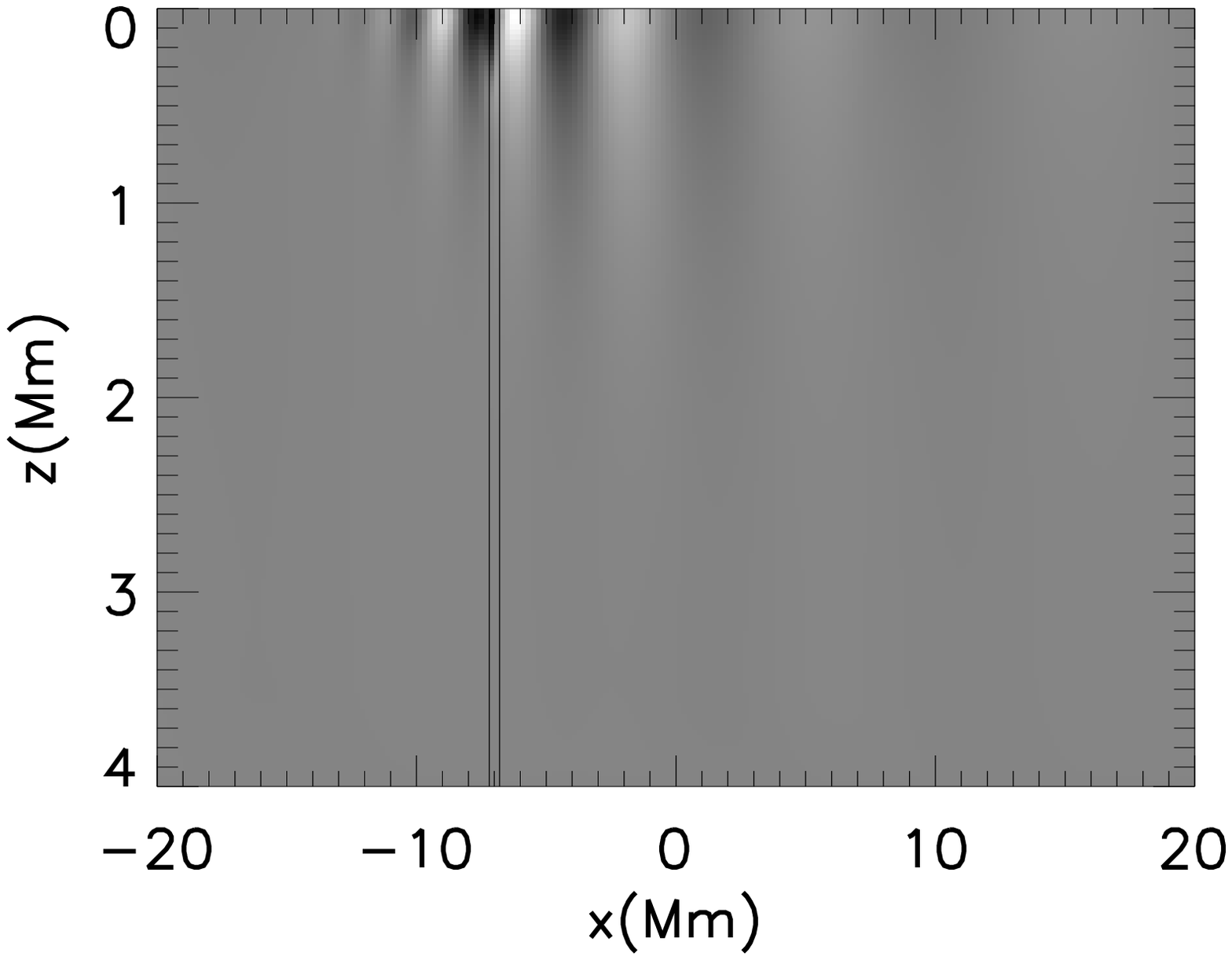}
\includegraphics[width=0.32\textwidth]{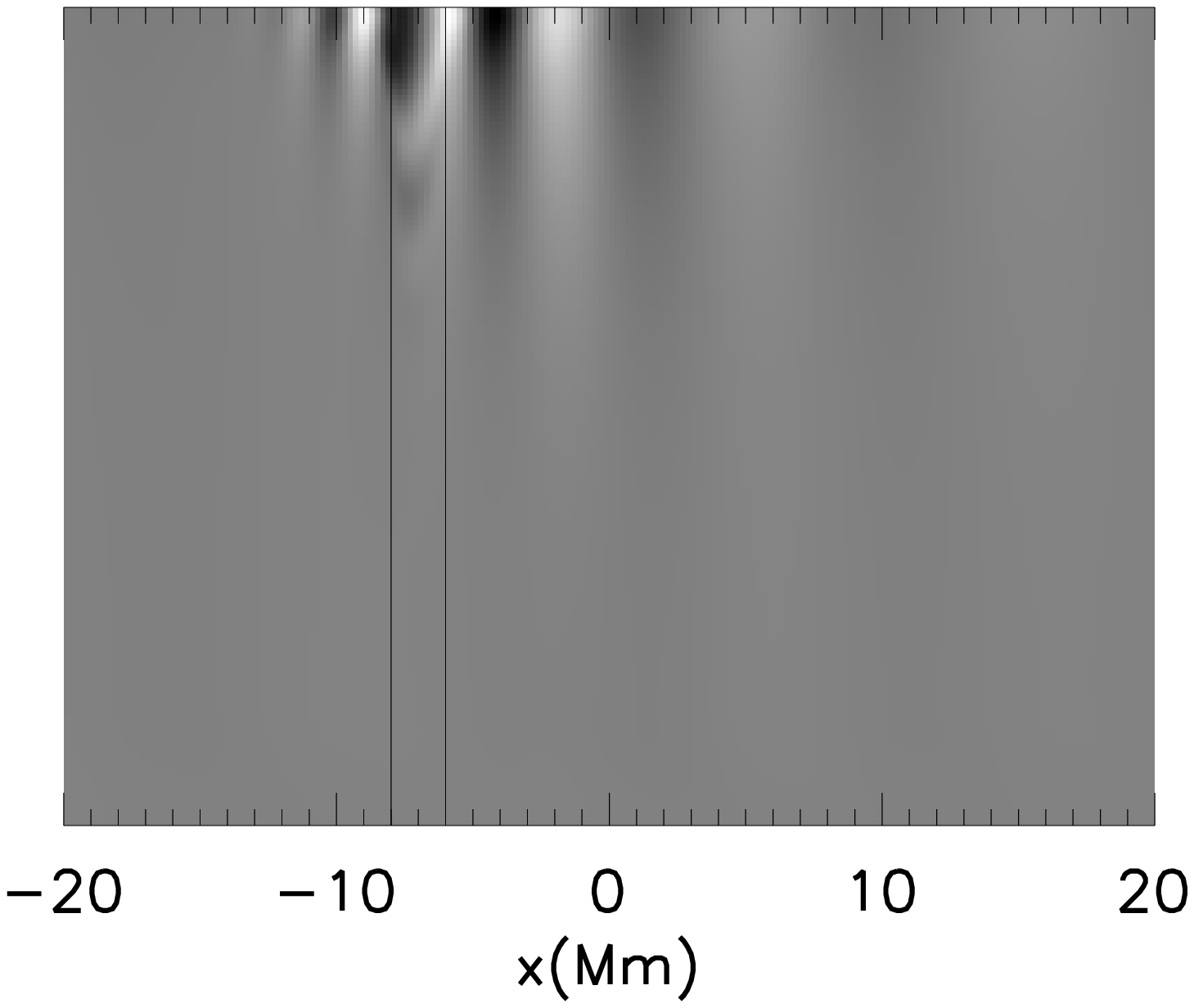}
\includegraphics[width=0.32\textwidth]{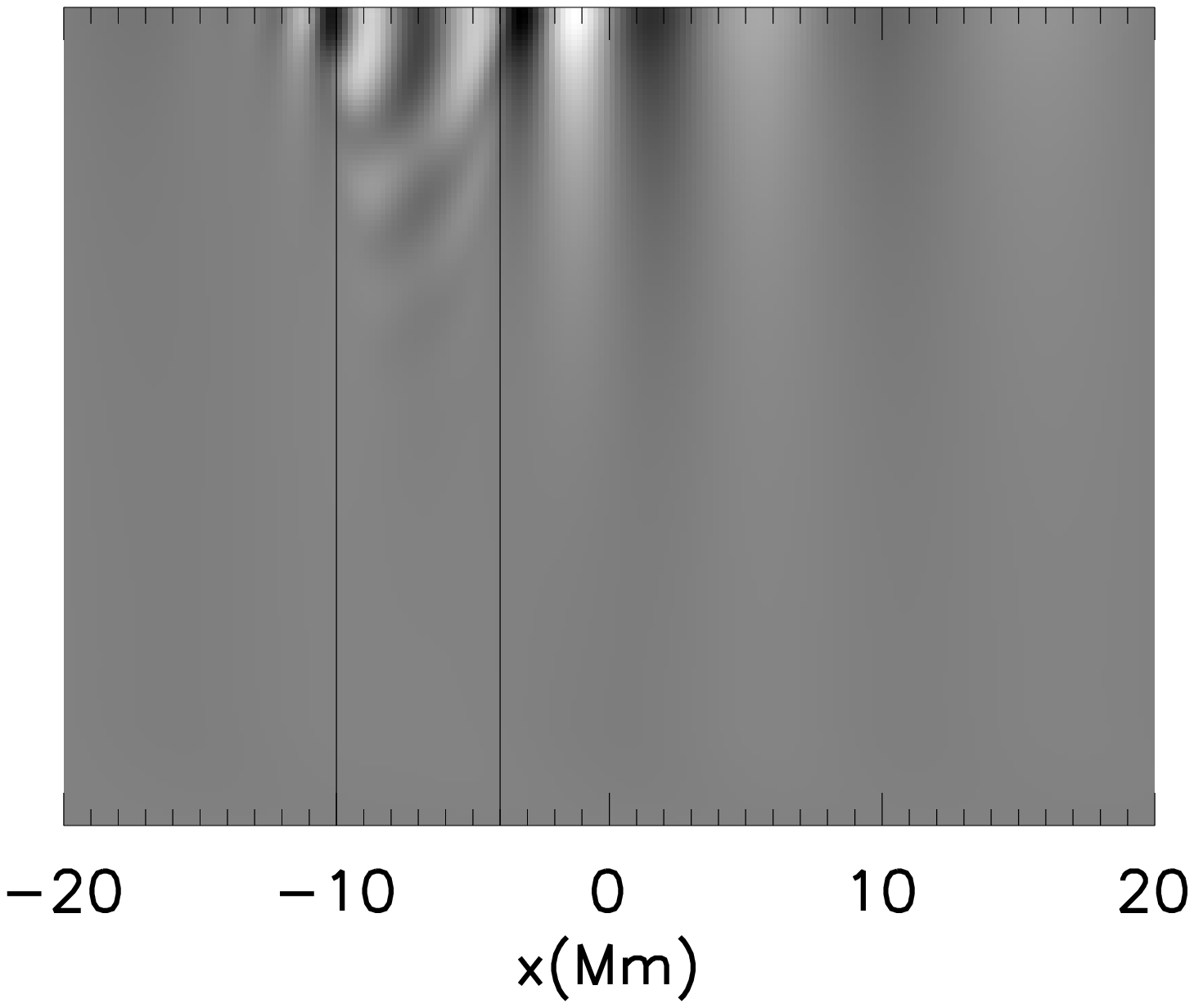}
\vspace{0.05\textwidth}
\caption{\ The $x$-component of the velocity near the photosphere; a vertical cut passing through the centre of the tubes 200 km, 1 Mm, 3 Mm. All panels are from a snapshot at $t$ = 2000 seconds.}
\label{3z}
\end{figure}

The dependence of the magnitude of the scattering on radius can be
seen in Figure~\ref{4rxy} where the full and the scattered wave field
are shown for tubes with radii 200 km, 500 km, 1 Mm, and 3 Mm. A vertical
cut through the spot, along the $x$-axis, is shown in Figure~\ref{3z}, for tubes with radii 200 km, 1 Mm, and 3Mm. For the two larger tubes, some mode conversion is apparent.

\begin{figure}    
\centering
\vspace{0.05\textwidth}
\includegraphics[width=0.36\textwidth]{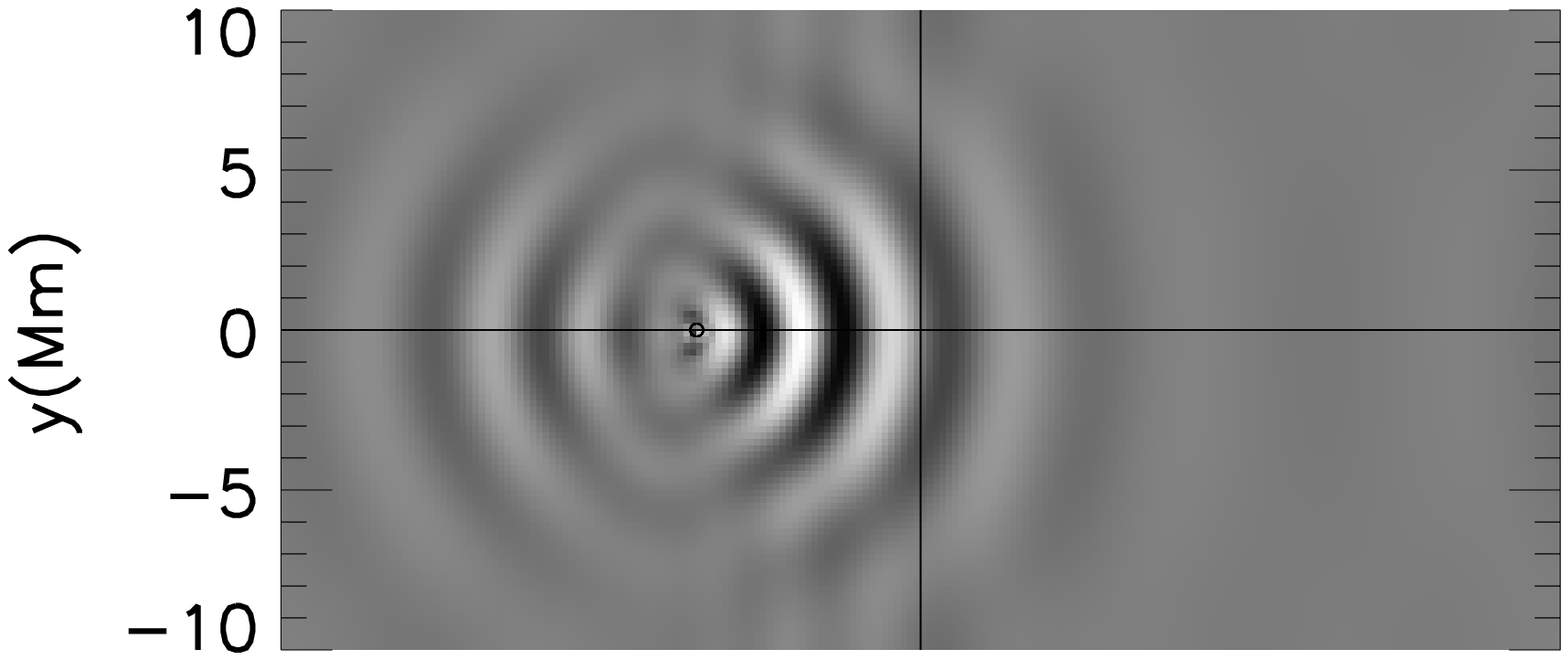}
\vspace{0.02\textwidth}
\includegraphics[width=0.31\textwidth]{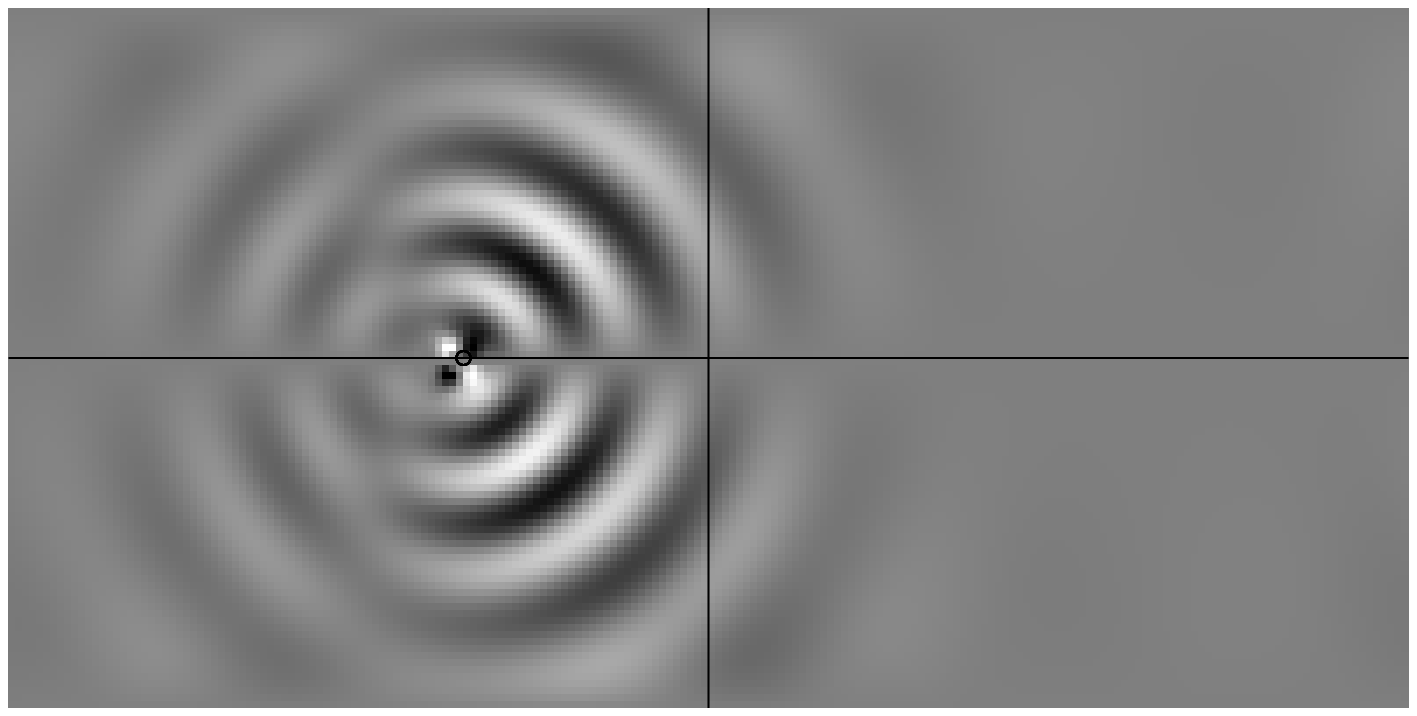}
\includegraphics[width=0.31\textwidth]{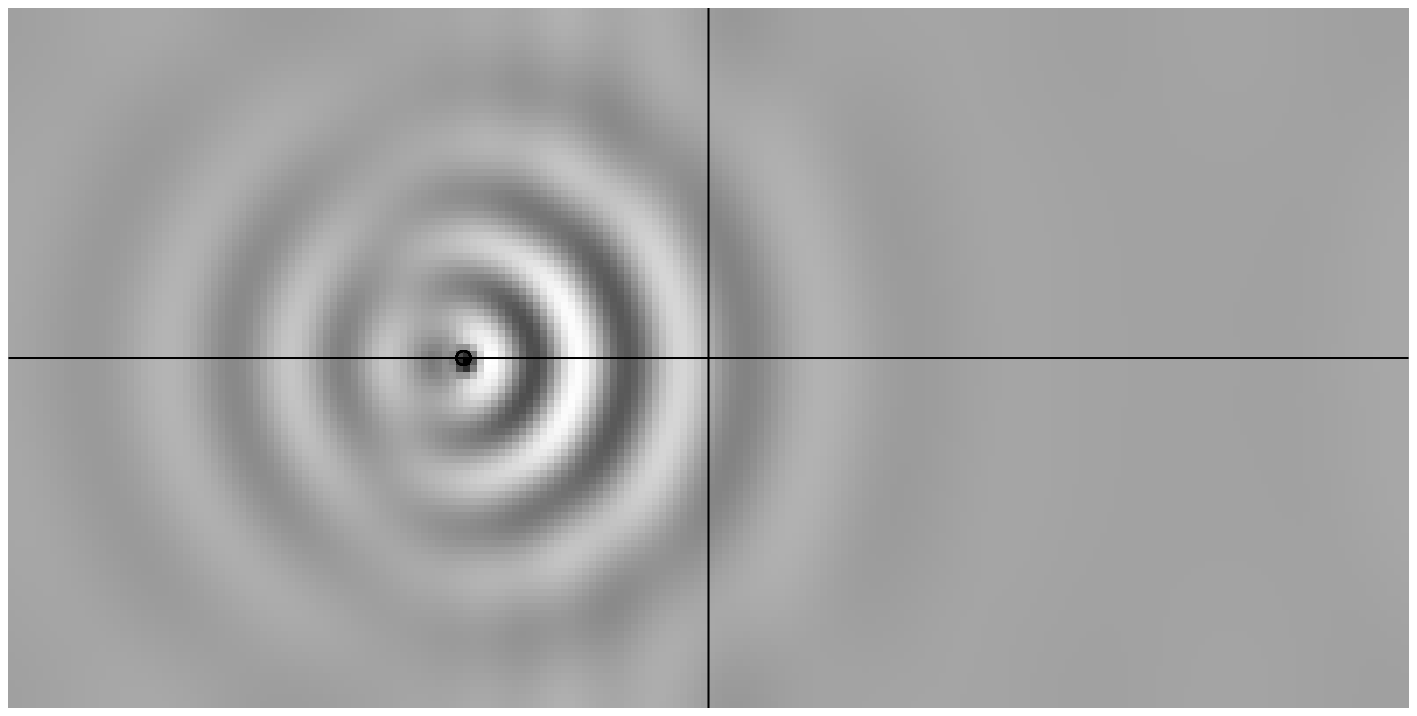}
\includegraphics[width=0.36\textwidth]{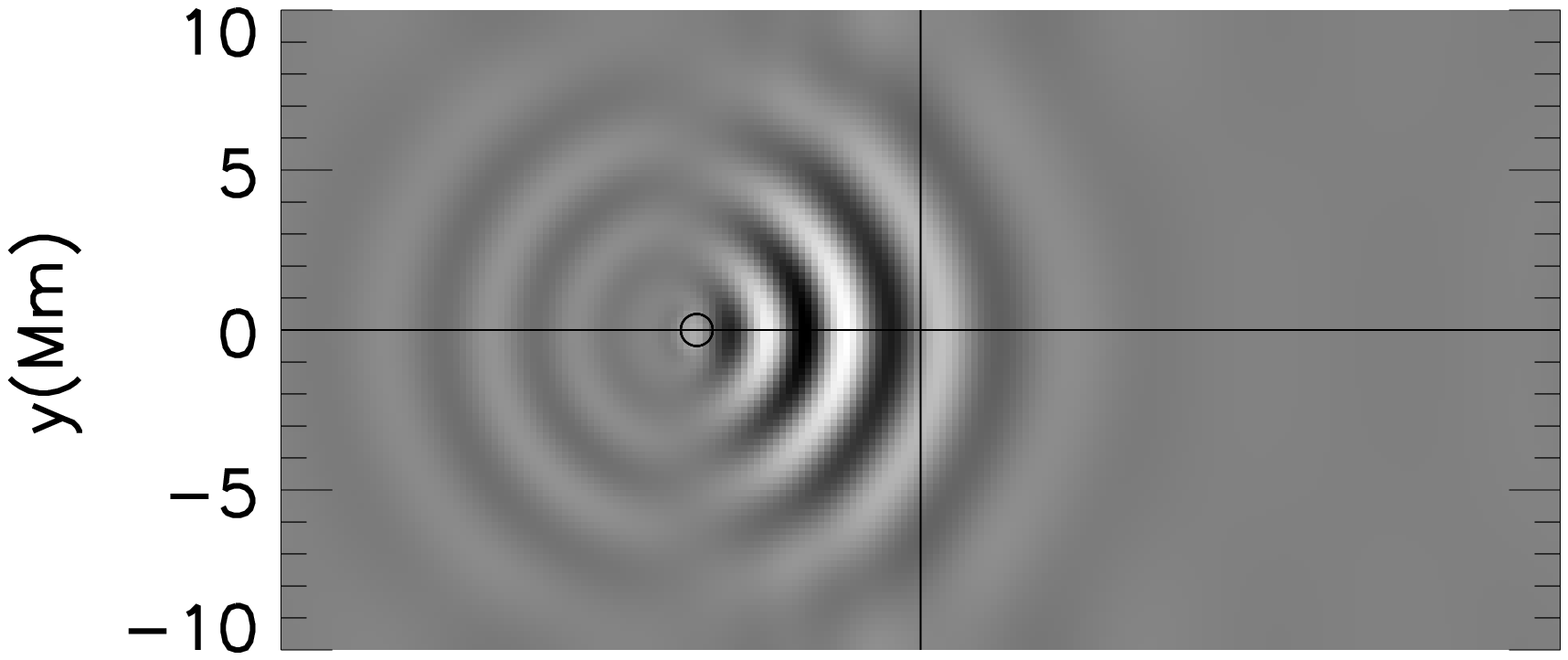}
\vspace{0.02\textwidth}
\includegraphics[width=0.31\textwidth]{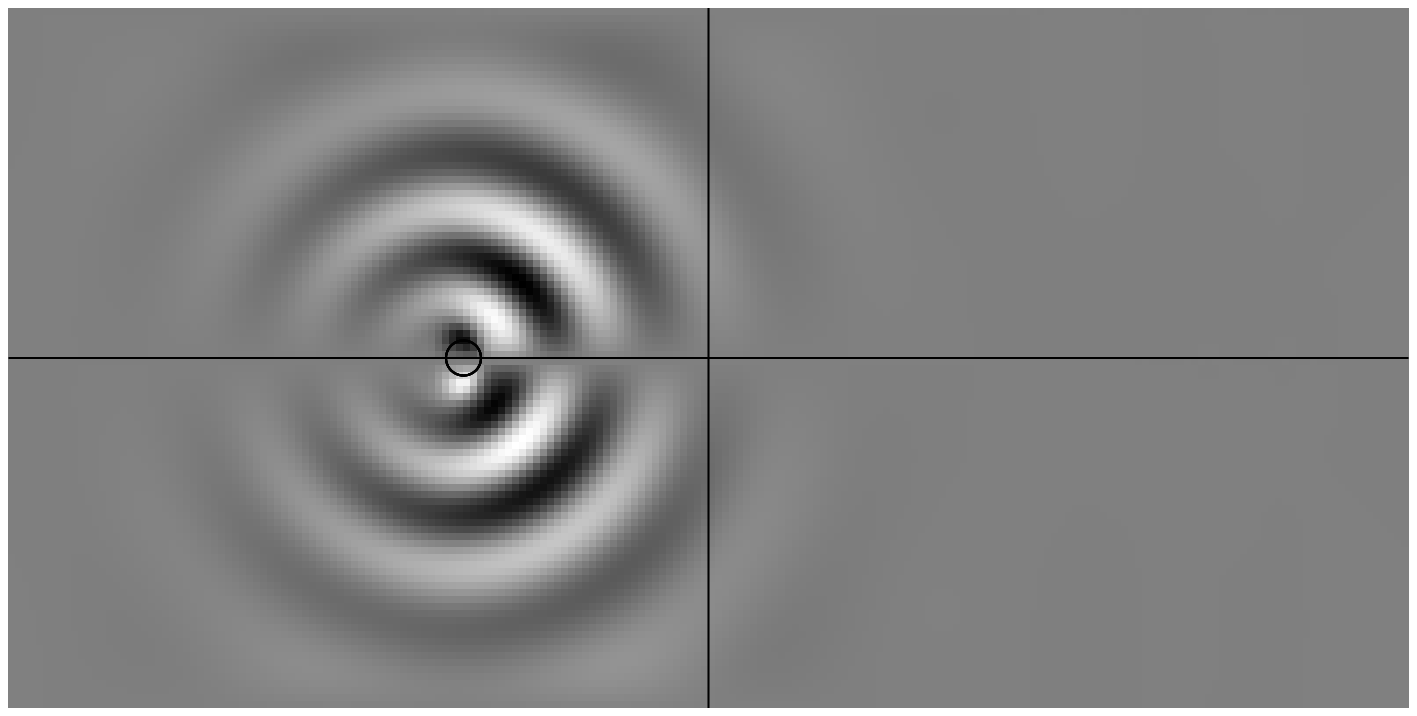}
\includegraphics[width=0.31\textwidth]{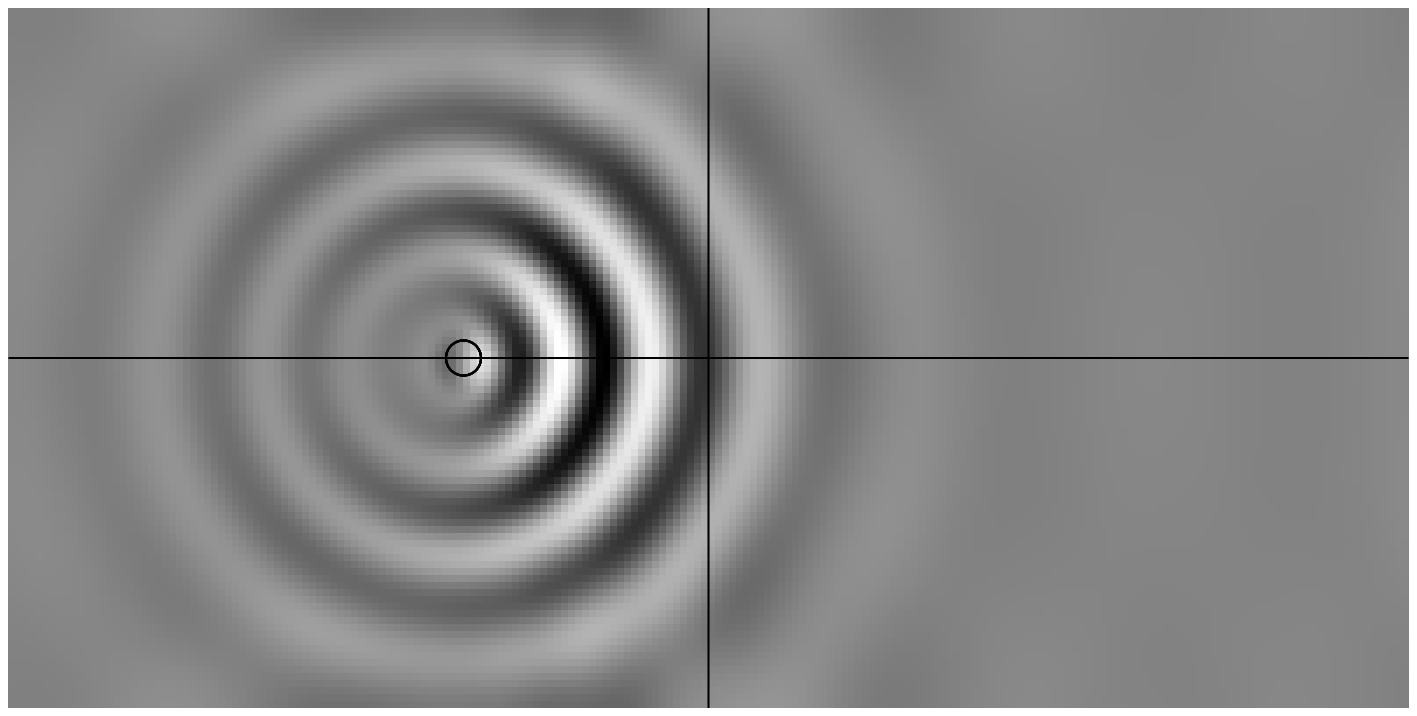}
\includegraphics[width=0.36\textwidth]{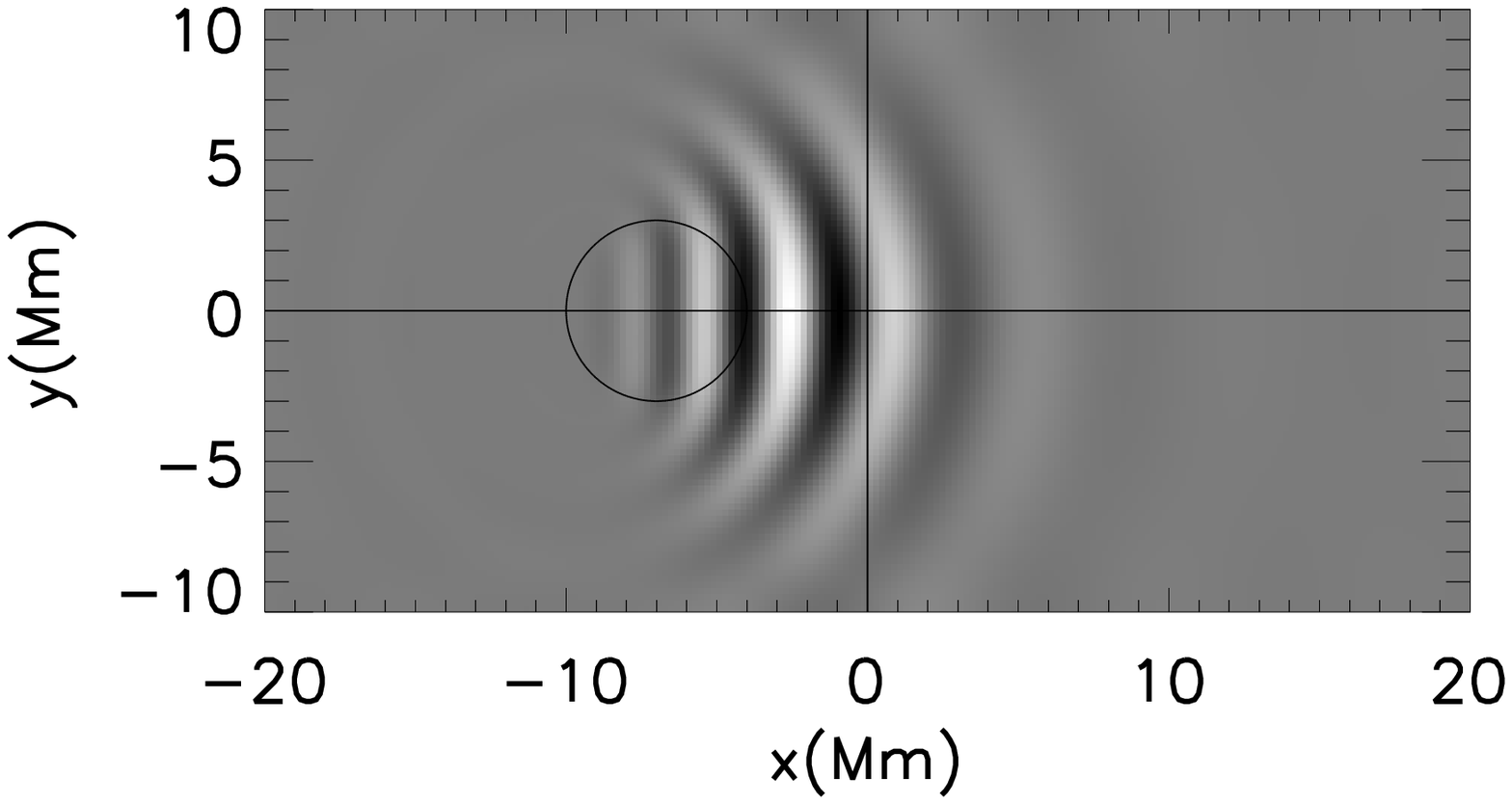}
\includegraphics[width=0.31\textwidth]{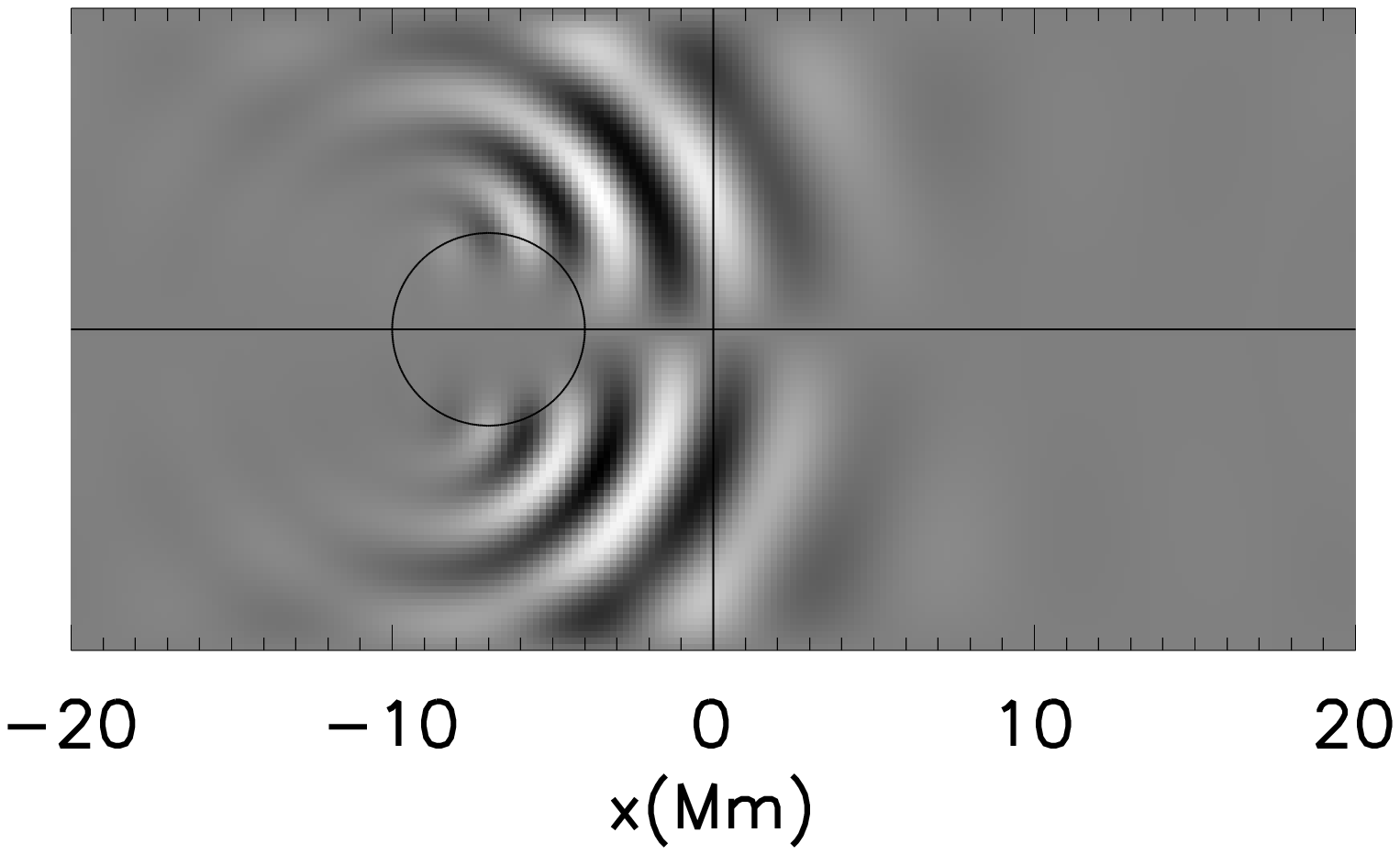}
\includegraphics[width=0.31\textwidth]{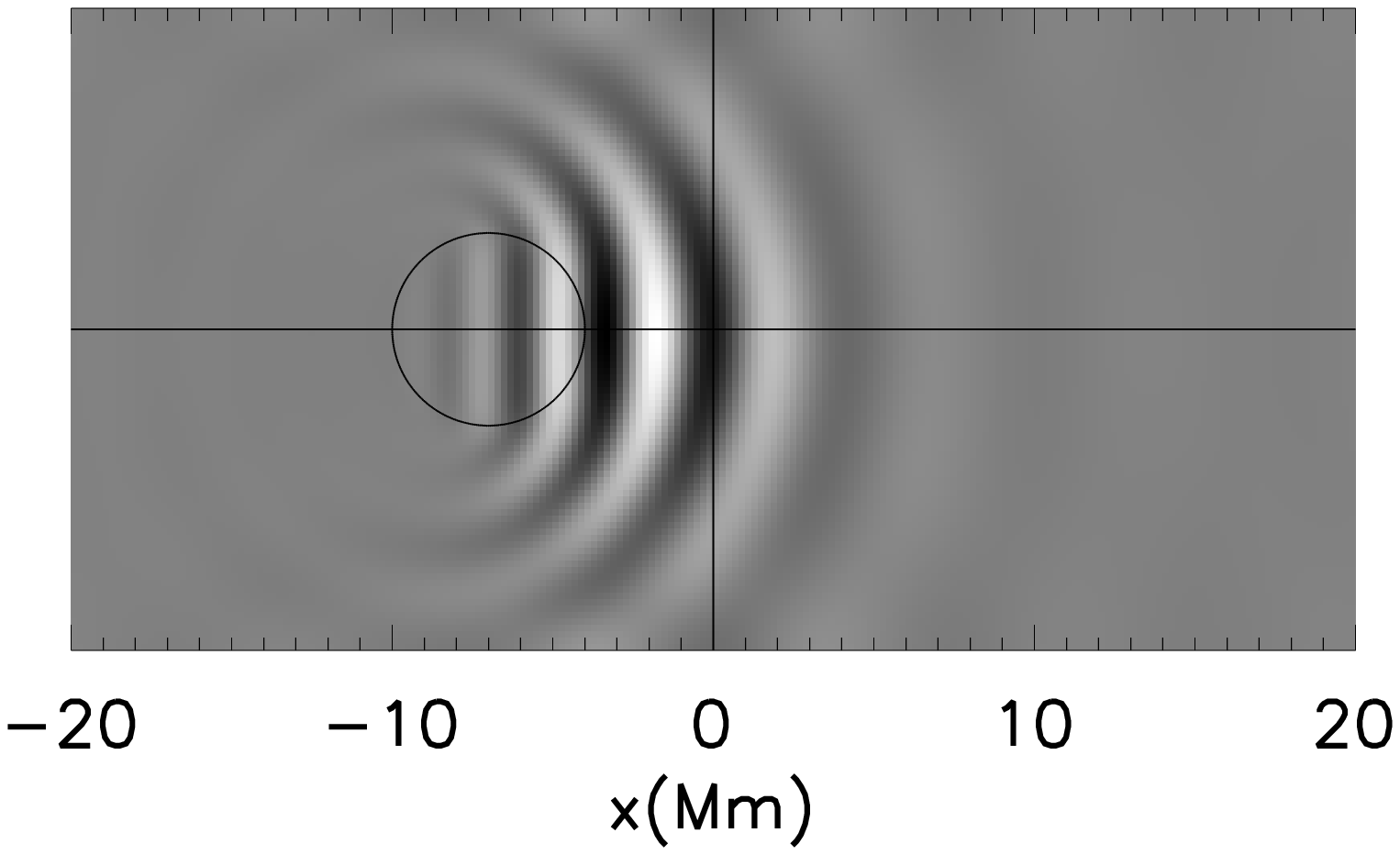}

\vspace{-0.63\textwidth}   
     \centerline{\bf      
      \hspace{0.13\textwidth}  \color{black}{$(V_{x})$}
      \hspace{0.26\textwidth}  \color{black}{$(V_{y})$}
      \hspace{0.25\textwidth}  \color{black}{$(V_{z})$}
         \hfill}
     \vspace{0.6\textwidth}

\caption{  Scattered wave field for different tubes
  radii taken 3000 seconds after the start of the simulation. Radius
  from top to bottom: 200 km, 500 km, 3 Mm. Velocity component
  from left to right: $V_{x}$, $V_{y}$, $V_{z}$. Note that different
  gray-scales are used for each frame so that only the structure
  of the scattered wavefield is significant}
\label{9xyvall}
\end{figure}

Naturally the simulations give us the full vector-velocity field and hence we can look simultaneously at all three components of the velocity field as shown in Figure~\ref{9xyvall}. This figure makes it clear that the scattered  field consists mainly of different mixtures of $m=0$ and $m=1$ modes for different radii.
From this figure we can see that tubes with radius between 500 km and 2 Mm are excited primarily in the  $m$ = 0 mode, the upper part of the cylinder moving inward or outward (sausage modes). For the $z$-component of velocity, tubes with radii less than 500 km are excited with $m = \pm$1 dipole oscillation whereas tubes between 500 km and 2 Mm oscillate with a mixture of $m=0$ and $m=1$ modes. It is more difficult to identify the oscillation of 3 Mm tube radius,
which presumably has substantial energy in higher $m$ components.

\begin{figure}    
\centering
\includegraphics[width=0.51\textwidth]{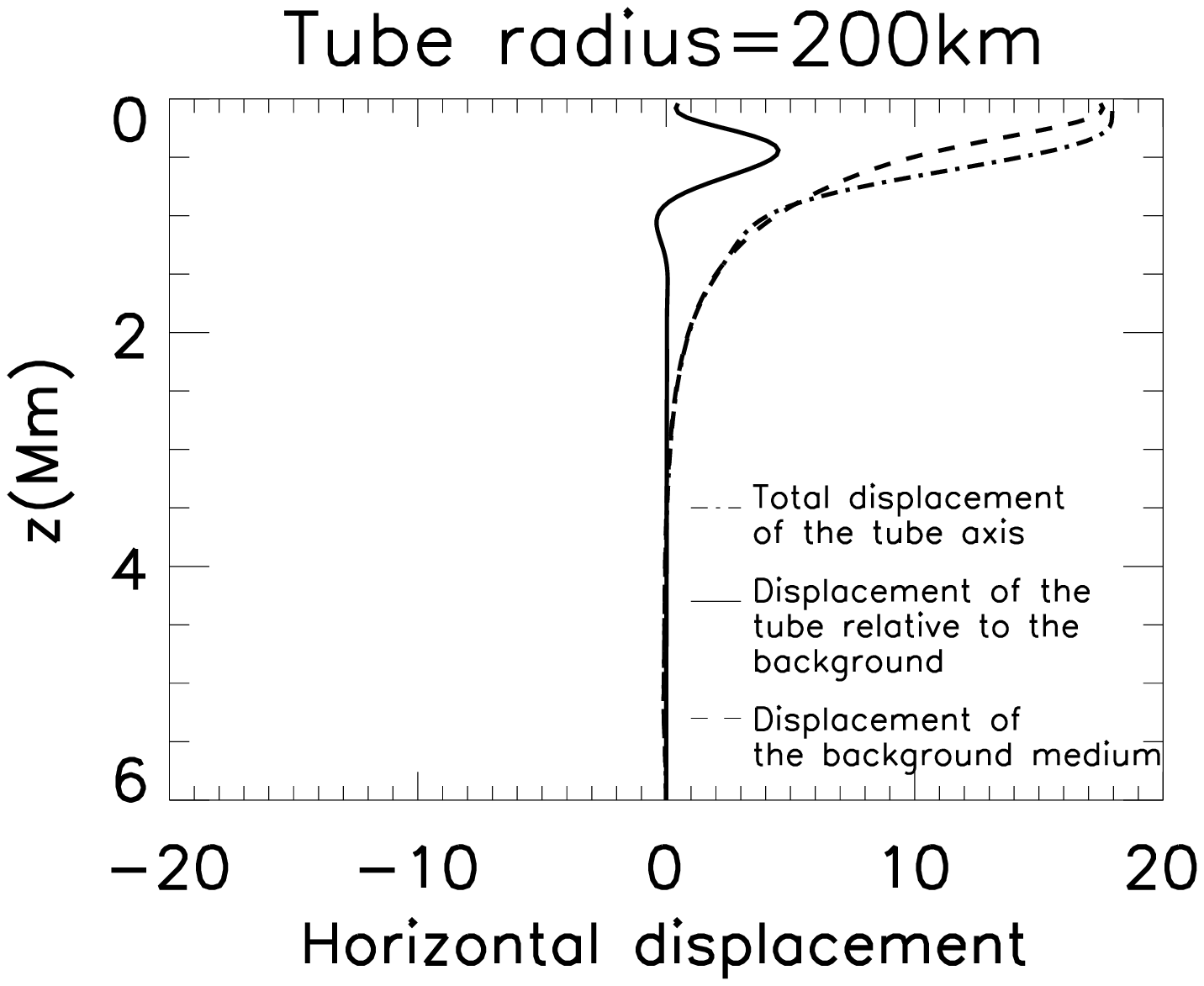}
\includegraphics[width=0.48\textwidth]{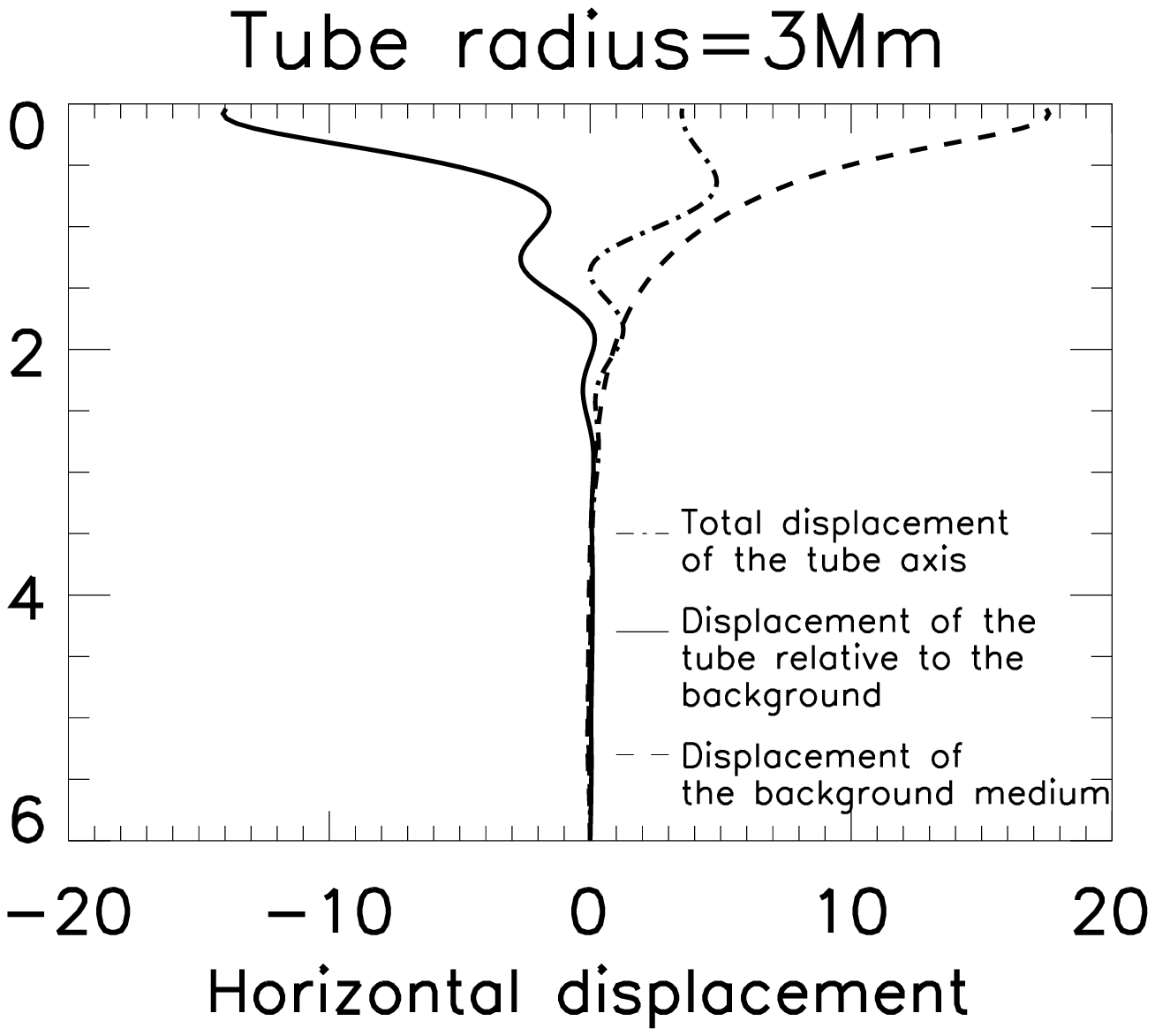}
\caption{\  Horizontal displacement in arbitrary units in the $x$-direction of 
  the tube axis and the
  background as a function of the depth [$z$].  The curves are taken
  2000 seconds after the start of the simulation for 200 km tube
  radius, left, and 3 Mm tube radius, right.}
\label{2rxz}
\end{figure}

\begin{figure}    
\centering
\includegraphics[width=0.505\textwidth]{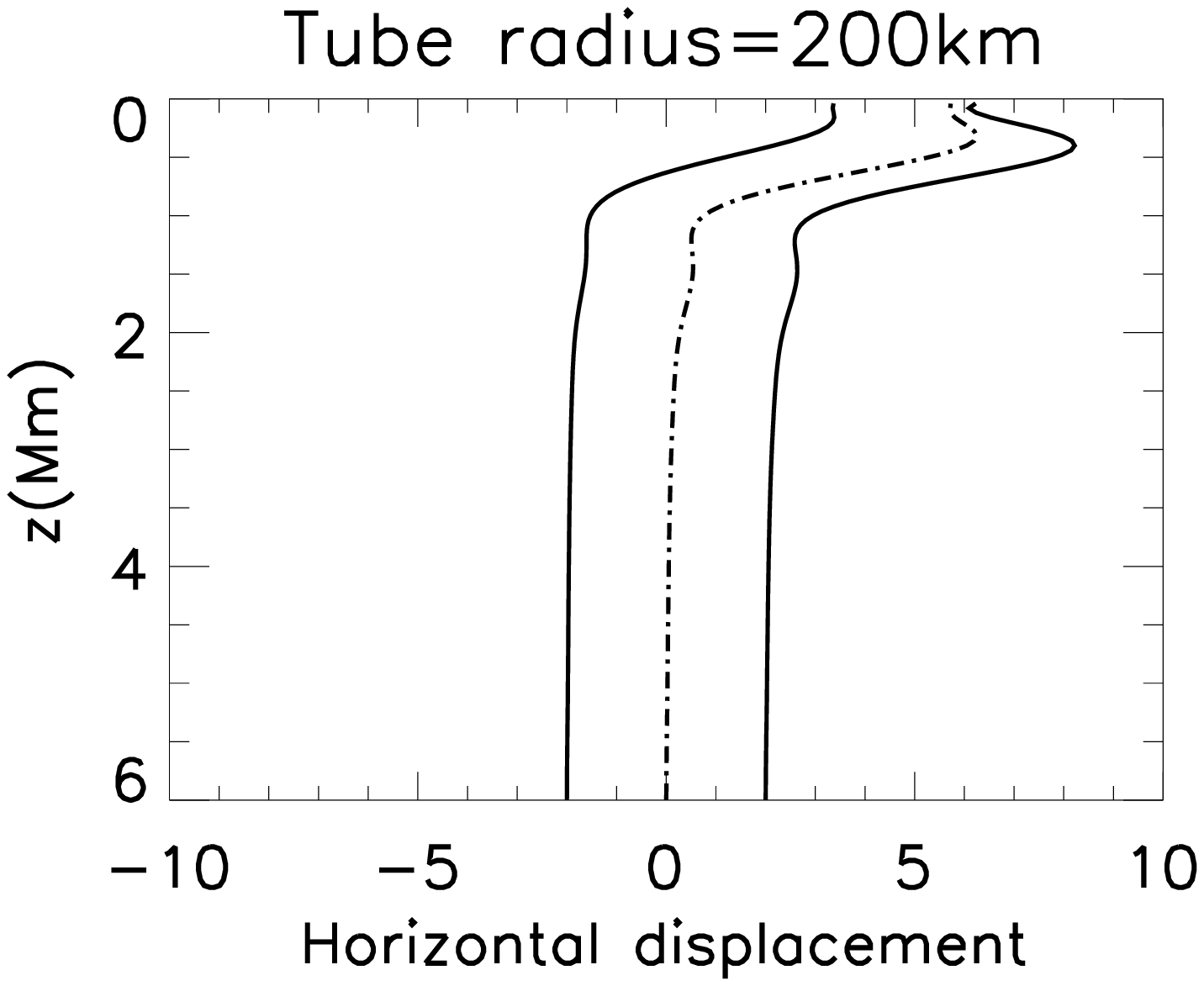}
\vspace{0.01\textwidth}
\includegraphics[width=0.48\textwidth]{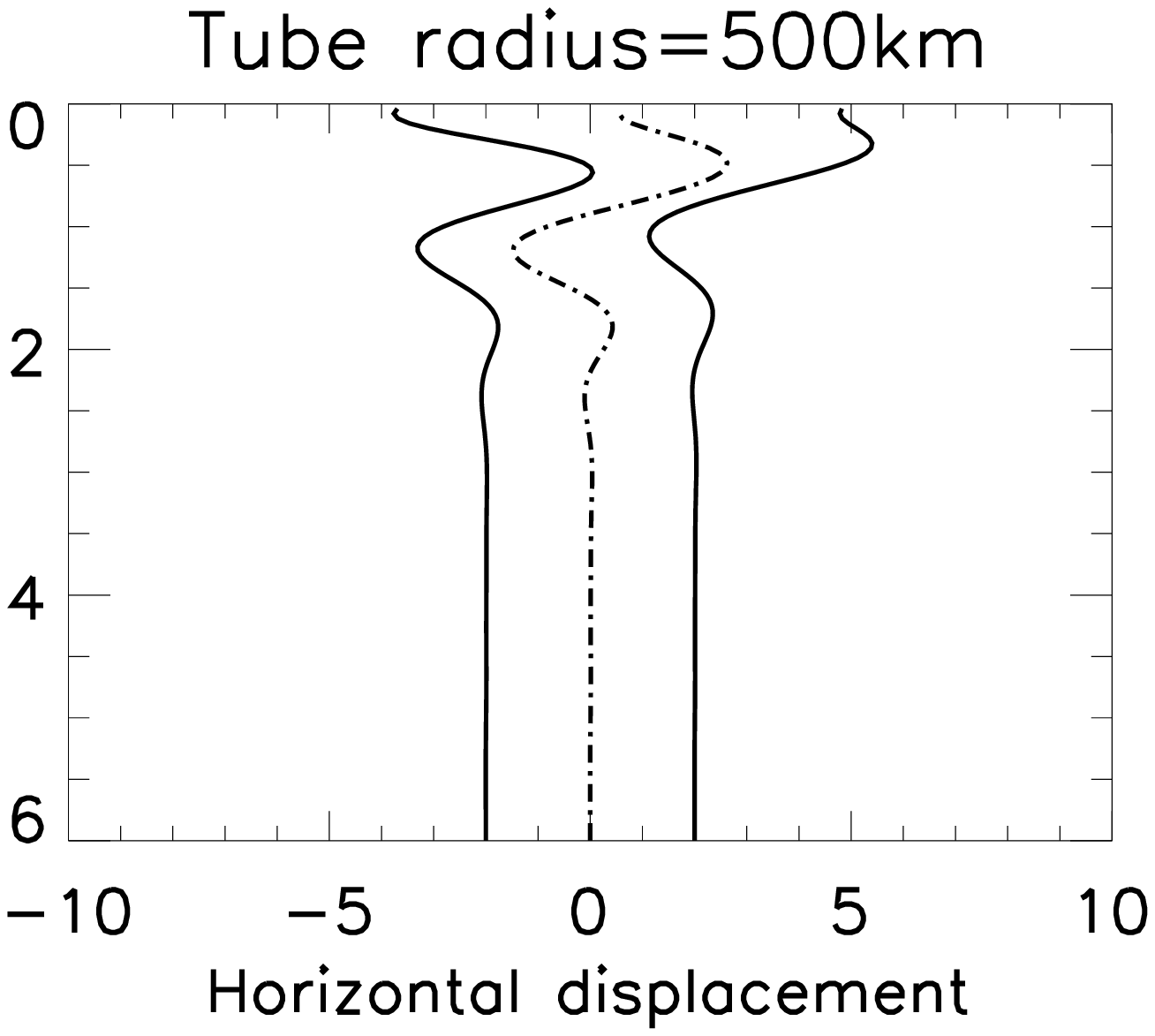}
\includegraphics[width=0.505\textwidth]{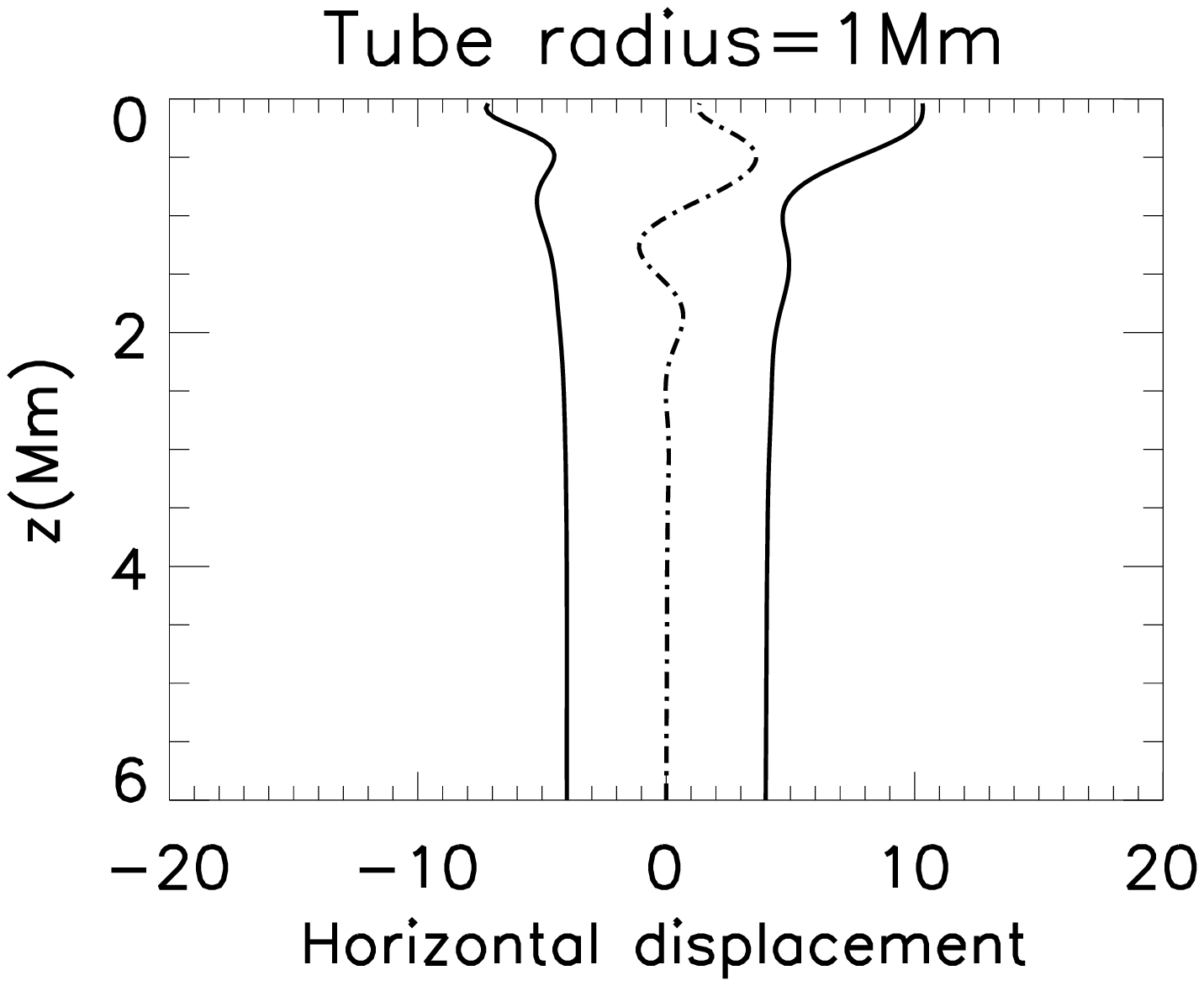}
\includegraphics[width=0.48\textwidth]{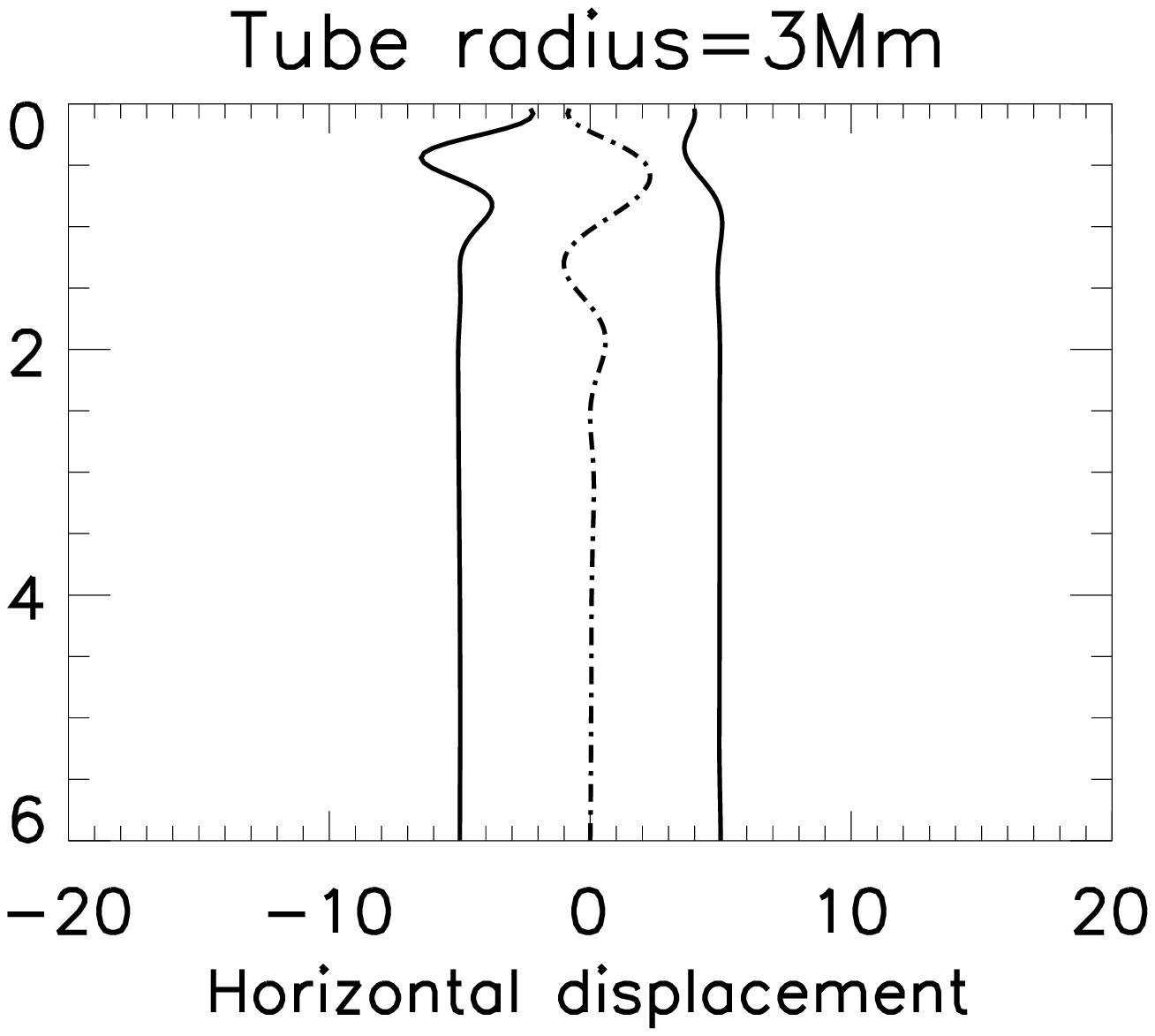}
\caption{\  Horizontal displacement in arbitrary units in the $x$-direction 
  of the tube axis (dashed
  line at the center) and the tube cross sections (full line right and
  left) as a function of the depth [$z$]. The curves are taken
  1600 seconds after the start of the simulation for different tubes
  radii: 200 km, 500 km, 1 Mm, and 3 Mm respectively.}
\label{4xzrl}
\end{figure}

To procede further we consider the displacement of the tube with
respect to its surroundings. The idea is illustrated in
Figure~\ref{2rxz}: Thinner flux tubes (radius smaller than 500 km) are
effectively coupled to the motion of the surrounding plasma by the
drag force, while larger tubes (radius larger than 500 km) can move relative to the surrounding plasma owing to the action of the magnetic force, similar to a flexible solid body immersed in a fluid \citep{Solanki06}. This result was expected, but is still nice to see in the simulations.

To better understand the $m$ modes that have been excited in the
tube, we look at the total $x$-displacement not only along axis of the
tube but also near its edges. This is shown in Figure~\ref{4xzrl}. Consistent with the above description, we see that the perturbations are coherent across small tubes. The change in behavior around 500 km corresponds to the fact that this is near the characteristic length-scale of the $f$ mode ($g/\omega^2$) of 771 km for the 3 mHz centered wave packet. For larger tubes, higher-$m$ modes begin to dominate because there is little coherence of the incoming wave across the tube.

\section{Mode Conversion}
\begin{figure}    
\centering
\includegraphics[width=0.48\textwidth]{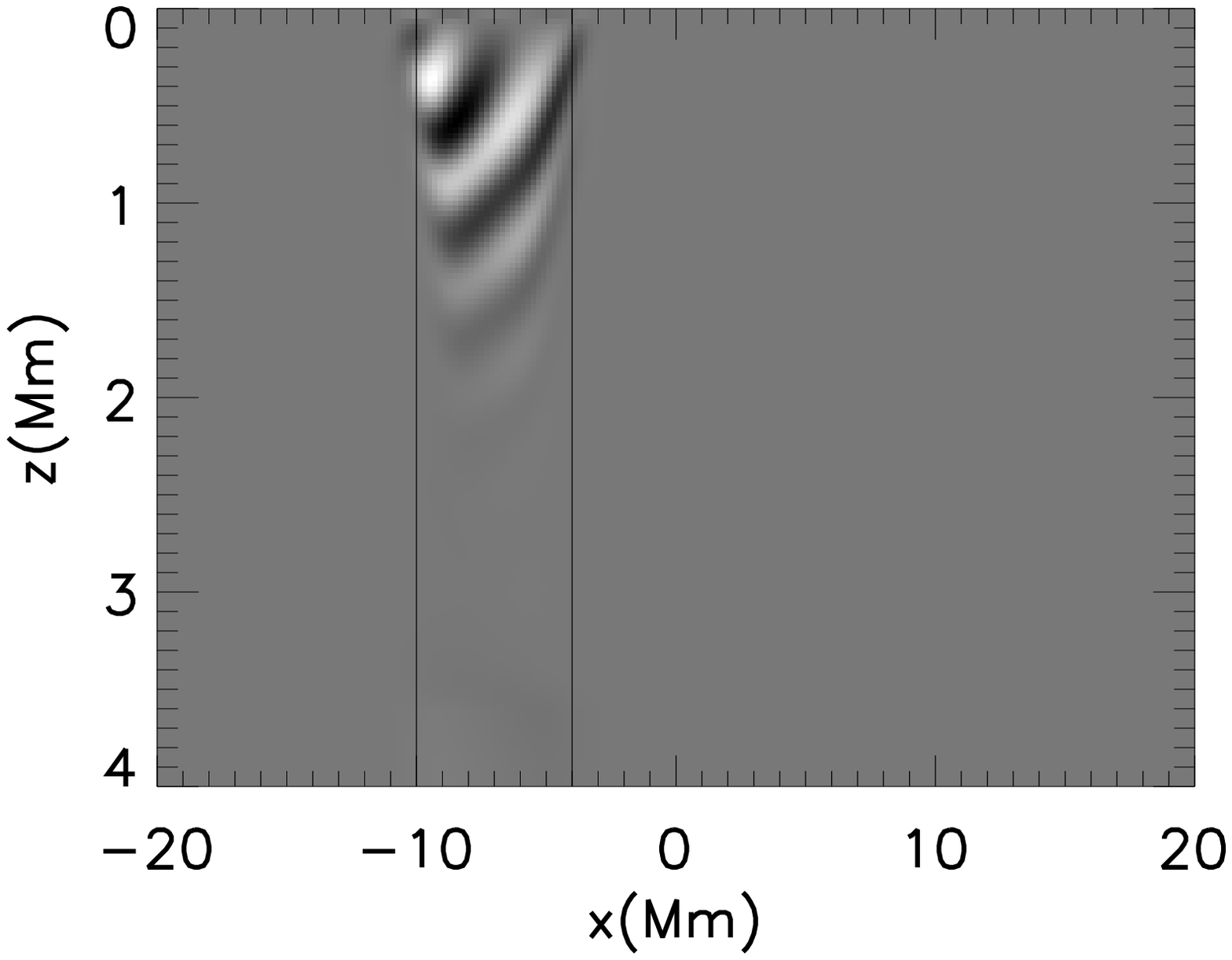}
\includegraphics[width=0.48\textwidth]{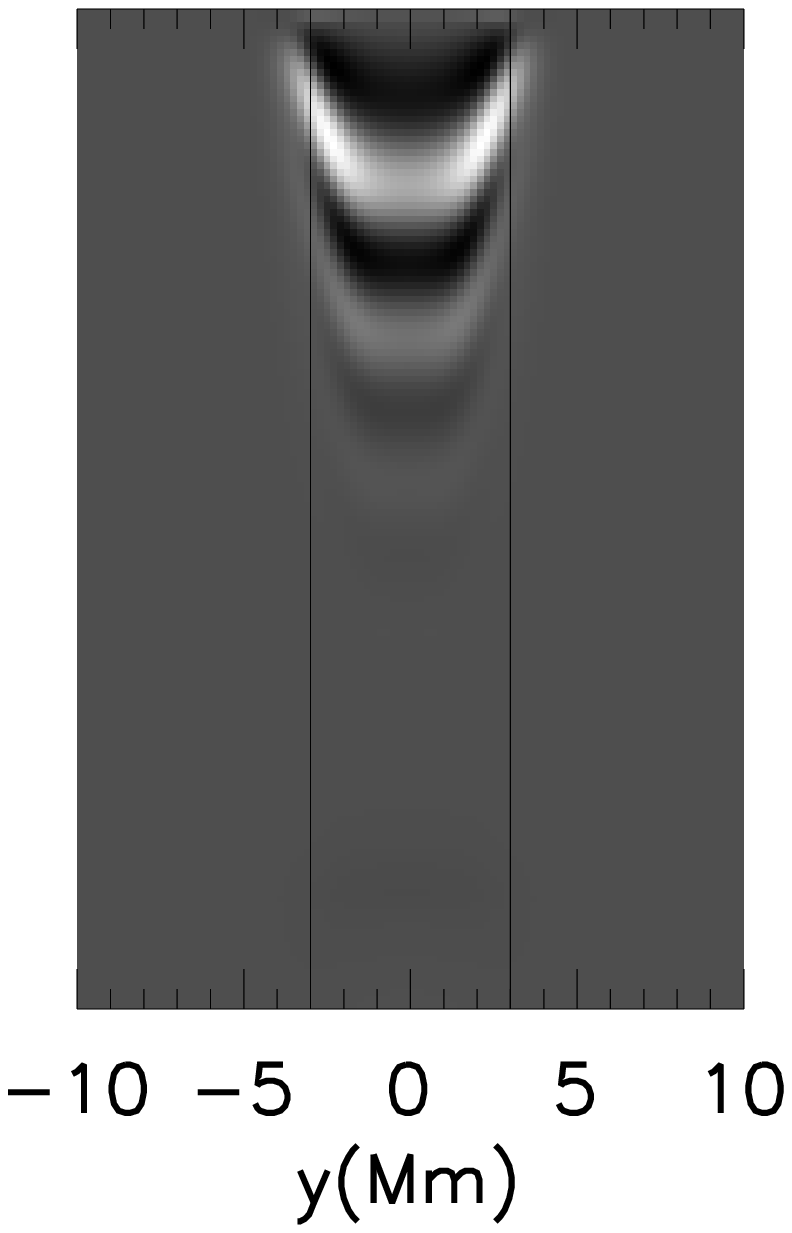}
\vspace{0.1\textwidth}
\caption{\  The $x$-component of the perturbation of the magnetic field
introduced by the passage of the wave, shown in the $x-z$ (left)
and $y-z$ (right) planes passing through the axis of the sunspot. The downward-propagating oscillations correspond to the slow-acoustic mode propagating along 
the fieldlines. This figure is for $t = 2280$~seconds and the 3~Mm flux tube.}
\label{f9}
\end{figure}

As in  \cite{Cally97}, we find that the magnetic flux tubes convert some of the incoming
waves into slow-magnetoacoustic modes which propagate along the field-lines. 
This can be seen in Figure~\ref{2rxz}, where the horizontal displacement of the flux
with respect to the background is shown. The vertical oscillations shown here are one
signature of downward propagating slow mode waves. The magnetic nature of these waves can be 
easily seen in Figure~\ref{f9} where a component of the magnetic-field perturbation is shown.
We comment that the mode conversion removes energy from the incoming waves. 

\section{Conclusions}
\label{S-Conclusions}
We have carried out a series of numerical simulations of plane-wave propagation through flux tubes of different sizes. In the first part of this study, we
tested the convergence in order to determine the minimum-sized tube that we could reliably simulate with the resolutions that we had available. We then investigated how different-sized tubes interacted with an incoming $f$-mode wavepacket. Different scattered wave-field patterns were observed for flux tubes with different radii. In agreement with previous suggestions, we found that when the flux tube was small compared with the wavelength of the incoming wave, mainly the $m=1$ kink modes were excited. For mid-ranged tubes both the $m=1$ kink and $m=0$ sausage modes were excited, and for large tubes numerous $m$ and radial order modes were excited. Our results demonstrate that numerical simulations are an efficient, robust, and straightforward method to treat the action of flux tubes on waves regardless of their size. The simulations are a rich source of information, which can be used to better understand the observations in future studies.


\bibliographystyle{spr-mp-sola}

\bibliography{SOLA_bibliography_example}

\IfFileExists{\jobname.bbl}{} {\typeout{}
\typeout{****************************************************}
\typeout{****************************************************}
\typeout{** Please run "bibtex \jobname" to obtain} \typeout{**
the bibliography and then re-run LaTeX} \typeout{** twice to fix
the references !}
\typeout{****************************************************}
\typeout{****************************************************}
\typeout{}}
\end{article} 
\end{document}